# Evolutionary Model of a Anonymous Consumer Durable Market

Joachim Kaldasch

EBC Hochschule Berlin

Alexanderplatz 1

10178 Berlin

Germany

(email: joachim.kaldasch@international-business-school.de)

**Abstract**

A analytic model is presented that considers the evolution of a market of durable goods. The model suggests that after introduction goods spread always according to Bass diffusion. However, this phase will be followed for durable consumer goods by a diffusion process governed by a variation-selection-reproduction mechanism and the growth dynamics can be described by a replicator equation.

The theory suggests that products play the role of species in biological evolutionary models. It implies that the evolution of man made products can be arranged into an evolutionary tree. The model also suggests that each product can be characterized by its product fitness. The fitness space contains elements of both sites of the market, supply and demand. The unit sales of products with a higher product fitness compared to the mean fitness increase. Durables with a constant fitness advantage replace other goods according to a logistic law. The model predicts in particular that the mean price exhibits an exponential decrease over a long time period for durable goods. The evolutionary diffusion process is directly related to this price decline and is governed by Gompertz equation. Therefore it is denoted as Gompertz diffusion.

Describing the aggregate sales as the sum of first, multiple and replacement purchase the product lifecycle can be derived. Replacement purchase causes periodic variations of the sales determined by the finite lifetime of the good (Juglar cycles). The model suggests that both, Bass- and Gompertz diffusion may contribute to the product life cycle of a consumer durable.

The theory contains the standard equilibrium view of a market as a special case. It depends on the time scale, whether an equilibrium or evolutionary description is more appropriate. The evolutionary framework is used to derive also the size, growth rate and price distribution of manufacturing business units. It predicts that the size distribution of the business units (products) is lognormal, while the growth rates exhibit a Laplace distribution. Large price deviations from the mean price are also governed by a Laplace distribution (fat tails). These results are in agreement with empirical findings. The explicit comparison of the time evolution of consumer durables with empirical investigations confirms the close relationship between price decline and Gompertz diffusion, while the product life cycle can be described qualitatively for a long time period.

**Keywords:** Evolutionary Economics, Economic Growth, Product Diffusion, Bass Diffusion, Gompertz Diffusion, Product Life Cycle, Price Distribution, Laplace Distribution, Replicator Equation, Firm Growth, Growth Rate Distribution,

## 1. Introduction

The evolutionary model presented here is based on the idea that durable goods produced for an anonymous market can be considered to be governed by the evolutionary Variation-Selection-Reproduction (VSR) mechanism, known form the evolution of species (for an overview [1,2]). For species the VSR-process can be summarized as follows:

Species reproduce by creating offspring (R). During this reproduction process random mutations lead to a variation of the individuals (V). Selection operates whenever different types of individuals reproduce at different rates (S). The frequency of the fittest offspring is increased. The VSR-process improves therefore the overall fitness of a species over a long time period.

We want to apply the idea of a preferential growth to durable goods. They are often produced for an anonymous market by different manufacturers. The goods differ slightly in their features. We want to denote these variants as models or products. Each model is manufactured by a business unit, while firms in general consist out of a number of business units. Business units manufacture products, sell them to the market in order to manufacture more. For durable goods the VSR-mechanism can be considered as follows:

Business units reproduce the good (R). During the reproduction process they vary the features of the models (V). The preferences of potential consumers determine the selective "environment" and hence a part of the fitness space of a product. The fittest models have an evolutionary (competitive) advantage compared to other models. Their reproduction rate is increased due to an increased financial return (S). Since the models with the lowest fitness disappear form the market, the overall fitness of the good will increase over a long time period.

An instructive example was given by Beinhocker [3], for the evolution of shirts. Shirts are manufactured by business units all over the world. Consumers have the choice between many different designs and prices. Some of the shirts are purchased more than others. The business units take advantage from this response and preferentially reproduce the best sellers. However, they also give designers the order to vary the best sellers. In the next generation non-best sellers disappear, while a number of variations exist for the best selling models. As a result, after sufficient time, the shirt market contains essentially shirts with the highest consumer preference.

The evolutionary theory presented below is based on this VSR-mechanism. However, it is much easer to vary the price instead creating permanently new versions of the product. The focus of the paper is rather on the impact of the price and the reproduction process on the market evolution, and does not explicitly consider the evolution of new models. Applying statistical methods a theory is developed that describes qualitatively the evolution of a durable consumer market over a long time period. However, in order to understand this growth, we have to emphasize the difference between spreading and evolution.

Let us return first to a biological example. Suppose we consider a plant, stranded on an empty island. If it founds ideal conditions, the average plant density (plants per unit area) increases until its reproduction is limited by the size of the island. This process is a spreading process. It depends on the reproduction rate of the plant and the space left to populate. No adaptation of the plant occurs. Now suppose the same situation, while the island is not empty, but filled with other plant species. After stranding the spreading process starts until it saturates at a density that will be lower than in the case of an empty island. However, the plant has the opportunity to increase its population density by adapting to the biological environment formed by the other species. This can be done for example by symbiosis (cooperation) with other species or by replacing them. The process responsible for this adaptation process is evolution. Both processes, spreading and evolution lead to an increase of the plant density, while the spreading process is usually much faster than the evolutionary VSR- mechanism.

Let us apply this idea to a durable good manufactured for an anonymous market (e.g. TV sets, refrigerators, PC's etc.). The good plays the role of the species (plant) and the market corresponds to the island. The market penetration, which counts the number of adopters (users) of the product scaled by the target population, characterizes the evolution of the good. This penetration process is denoted in economy as market diffusion [4]. After introduction the information about the product spreads by the word-of-mouth effect. Previous adopters inform their neighbours in the social network about the utility of the product. Those potential consumers, who can afford the product, will purchase the good during this spreading process. A model that describes this spreading was established by Bass [5]. Therefore we want to denote the corresponding spreading process as Bass diffusion. The spreading process increases the market penetration. However, during this process the features of the good remain constant.

The spreading process saturates after a sufficiently long time. For a low introduction price all potential consumers can purchase the good, (which corresponds to the empty island case). However, expensive consumer durables are not affordable for all consumers and the market penetration will saturate after Bass diffusion at a lower level (which corresponds to the filled island case). In this case the market penetration can be increased by a slight decrease of the product price. Since this variation leads to an increased return for the business units, the reproduction rate of the corresponding models is increased. The theory presented below suggests that the fitness of a product is directly related to the market volume, which is a function of the mean price. Since an increase of the market penetration can be interpreted as a diffusion process, there must be an additional diffusion process associated to the VSR-mechanism. In the presented evolutionary model the corresponding diffusion process can be described by Gompertz equation and is therefore denoted as Gompertz diffusion. Gompertz equation was originally developed in 1825 by the British actuary, Benjamin Gompertz, to describe the relationship between age and mortality [6,7]. The cumulative adoption curve is S-shaped, but unlike Bass penetration, the Gompertz curve is not symmetric. The highest rate of adoption occurs when around 37% of all eventual adopters have adopted.

A consequence of this evolutionary model is that both Bass and Gompertz diffusion may occur during the diffusion of an expensive consumer durable, while Bass diffusion takes place much faster. Gompertz diffusion must be directly related to a decreasing market price. As will be shown below, this prediction can be confirmed by empirical data.

Applying statistical methods the theory can be extended to establish a size and growth rate distribution of the business units (products), where the size is characterized by the unit sales. Also derived is the price distribution and mean price evolution of the models. This allows a specification of the unit sales over a long time period, known as the product life cycle.

The paper is organized as follows. The next section is devoted to the presentation of the model, while key assumptions are numbered by lower case roman letters. In order to show the applicability of the model a comparison with empirical investigations is performed in the third section of the paper, followed by a conclusion.

## 2. The Model

### 2.1. The Market Volume

We want to consider a consumer durable good that is made for an anonymous market. The market consists of an ensemble of agents who are interested in purchasing the good, denoted as market potential *M*.

It is an empirical fact that the annual income distribution, $P_I(h)$ shows a two-class structure [8]. The upper class can be described by a Pareto power-law distribution. The majority of the population, however, belongs to the lower class. We want to separate the market potential into an upper and lower class contribution:

$$M = M_L + M_U$$
(1)

where the relative market potential in the USA is $m_U = M_U/M \approx 1\text{–}3\%$ and $m_L = M_L/M \approx 97\%\text{–}99\%$ of the population [8].

The market volume *V(p)*, depends on the product price *p*. It is limited to those agents who have sufficient personal annual income, *h*, to afford the good. The market volume determines therefore the number of potential first purchase consumers for a given product price, denoted in this paper as potential adopters. In order to determine *V(p)* we make two general assumptions:

i) The upper class can always afford the good and is not limited by the product price.

Hence the market volume consists of the upper class and a part from the lower class $V_L(p)$, which depends on the product price:

$$V(p) = V_L(p) + M_U$$
(2)

As to evaluate *V(p)*, we take advantage form the income distribution of the lower class. This distribution can be described by an exponential Boltzmann-Gibbs distribution, a lognormal distribution or a *Γ*-distribution (except for zero income) with an appropriate choice of the free parameters [9,10,11]. For mathematical simplicity we confine here to the case that the income distribution of the lower class can be approximated by a Boltzmann-Gibbs distribution. In this case the relative abundance to find a representative agent having an annual income between *h* and *h+dh*, can be given by the probability density function (pdf):

$$P_I(h) = \frac{1}{I}\exp(-h/I)$$
(3)

where the average personal income can be obtained from:

$$I = \int_0^\infty dh P_I(h) h$$
(4)

ii) The key idea to evaluate the market volume is to assume that the chance to find an agent of the lower class who is willing to purchase the good, has a maximum at a minimum price, $p_m$.

The minimum price $p_m$ corresponds to the lowest willingness to pay. However, following standard microeconomics the willingness to pay is a function of the personal income. In order to take this effect into account, we scale the nominal product price $p$ by the mean income and introduce a real price:

$$\mu = \frac{p}{I}$$
(5)

where the mean income is given by Eq.(4). For a consistent notation, the model is developed in terms of real prices.

The second assumption implies that for a product price $\mu \leq \mu_m$, the market volume must be equal to the market potential:

$$V(\mu_m) = M$$
(6)

while for $\mu > \mu_m$, the market volume must be a function of the price.

The market volume $V_L(\mu)$ is determined by the chance to find an agent who has sufficient income to afford the good. The lower class contribution to market volume can therefore be written as:

$$V_L(\mu) = M_L \int_{z(\mu)}^{\infty} P_I(z')dz'$$
(7)

where $z' = h/I$. The integral determines the probability to find an agent with sufficient income, while the unknown function $z(\mu)$ specifies how this probability varies as a function of the price.

The function $z(\mu)$ can be derived from two conditions.

1. Because the cumulative income distribution is normalized to one, the function $z(\mu)$ must be zero at $\mu_m$, in order to fulfil Eq.(6). Hence, we can approximate the function $z(\mu)$ close to $\mu_m$ by a Taylor expansion. Up to the second order we can write for $\mu > \mu_m$:

$$z(\mu) \cong z_1(\mu - \mu_m) + z_2(\mu - \mu_m)^2$$
(8)

where $z_1, z_2 \geq 0$.

2. We demanded in *ii)* that the chance to find a potential consumer as a function of the price has a maximum at $\mu=\mu_m$. Hence, $dV_L(\mu_m)/d\mu=0$. This condition implies that $z(\mu)$ has a minimum at $\mu_m$, and hence, $z_1=0$.

The market volume can therefore be approximated near $\mu_m$ by:

$$V_L(\mu) \cong M_L \exp\left(-\frac{(\mu-\mu_m)^2}{2\Theta^2}\right)$$

(9)

where

$$\Theta^2 = \frac{1}{2z_2}$$

(10)

For later use we further introduce the market volume scaled by the market potential. The corresponding density is given by:

$$v(\mu) = \frac{V(\mu)}{M} = v_L(\mu) + m_U$$

(11)

with

$$v_L(\mu) = \frac{V_L(\mu)}{M}$$

(12)

For the case that the durable good has also industrial applications, we consider firms as agents not limited by the price, contributing to $M_U$. Since the market potential $M$ is a large figure, densities can be treated as continuous variables.

## 2.2. The Reproduction Process

In this chapter we want to study the reproduction process of a consumer good. The good is produced for an anonymous market and distributed either directly or by a retail system. For a given consumer good there exists a number of products (models) having similar utility properties. We want to denote these models and the corresponding business units with the index, $i$, where $N$ is the total number of models (business units). Note that firms often consist of a number of business units.

The reproduction process is a cycle as displayed schematically in Fig.1. Business units sell their products in order to manufacture new ones. They are treated as independent, such that they can vary the product price $\mu_i$ and supply (absolute output $S_i$) independently. The absolute number of sold models per unit time of the *i-th* business unit is $Y_i$. In order to establish a continuous theory we introduce the following densities for the purchase and supply flow:

$$y_i = \frac{Y_i}{M}; s_i = \frac{S_i}{M}$$
(13)

and the total unit sales and total supply of the durables is given by:

$$y_t = \sum_{i=1}^{N} y_i; s_t = \sum_i s_i$$
(14)

For later use we also introduce a mean price as the average over the sold products:

$$\langle \mu \rangle = \frac{1}{y_t} \sum_{i=1}^{N} y_i \mu_i$$
(15)

iii) The key idea to understand the dynamics of the reproduction process is the assumption that the business units try to compensate sales fluctuations. Small sales fluctuations can be balanced by means of inventories or small capacity adaptations. However, on large sales fluctuations they have to additionally respond by price adaptations. The reason for this assumption is that large variations of the production capacities require time consuming investments.

This assumption suggests that the variation of the total capacities is a function of the considered time scale. On a short term time scale (of the order of month) the total capacities of the market are nearly constant, while on a long term time scale (of the order of years) they can be varied considerably. Taking advantage from this constraint we introduce a separation of the time scales. The parameter $\tau$ indicates the short time scale and the parameter $t$ the long time scale, obeying the relation:

$$\tau = \varepsilon t$$
(16)

with $\varepsilon << 1$ .

Assumption iii) implies that the business units can respond immediately on small demand fluctuations according to:

$$\frac{ds_i(\tau)}{d\tau} = \gamma'_i \frac{dy_i(\tau)}{d\tau}$$
(17)

where the proportionality constant $\gamma'_i$ is denoted as the reproduction coefficient of a business unit. The short time scale is chosen such that the total supply is nearly constant. Hence:

$$\frac{ds_t}{d\tau} = \frac{dy_t}{d\tau} \cong 0$$
(18)

where we used Eq. (17).

The business units can be viewed as input-output systems (Fig.1). The input is the financial flow (revenue), which is the product of the number of sold models and its price. If there is no input flow $y_i$ the output flow $s_i$ will be zero. Therefore, Eq.(17) can be integrated, while the integration constant is zero. Hence:

$$s_i(\tau) = \gamma'_i\, y_i(\tau)$$
(19)

and with Eq.(14)

$$s_t = \sum_i \gamma'_i\, y_i = \langle \gamma' \rangle y_t$$
(20)

where $<\gamma'>$ is the mean reproduction coefficient.

Further we want to determine the density of the *i-th* product available in the market, $x_i(\tau)$. For the *i-th* product this density is determined by the balance between supply and purchase flow:

$$\frac{dx_i(\tau)}{d\tau} = s_i(\tau) - y_i(\tau) = \gamma_i y_i(\tau)$$
(21)

where $\gamma_i=\gamma'_i-1$, also denoted as reproduction coefficient. The total density of available products at a given time step is determined by:

$$x_t(\tau) = \sum_{i=1} x_i(\tau)$$
(22)

The next step is to specify the product demand. It is determined by the chance that an arbitrary agent becomes a potential consumer. This can be due to either first purchase or repurchase decisions. The aggregate number of potential consumers at time step $\tau$ scaled by the market potential is $\psi(\tau)$. Hence, the density of potential consumers can be given by:

$$\psi(\tau) = \psi_f(\tau) + \psi_r(\tau)$$
(23)

while first purchase and repurchase decisions are indicated by $\psi_f$ and $\psi_r$.

The purchase process can be considered as a statistical event, that a potential consumer meets the *i-th* model and purchase it. The function $y_i$ is obviously zero if there are either no potential consumers $\psi$ or available products $x_i$. Up to the first order, $y_i$ must be therefore proportional to the product of both densities, potential consumers and available products. Hence, purchase events occur with a frequency:

$$y_i(\tau) \cong \eta_i x_i(\tau)\psi(\tau)$$
(24)

where the rate $\eta_i$>0 characterizes the mean success of the *i-th* model and is denoted as the preference parameter. It is essentially determined by the product utility and the spatial distribution of the product. The purchase process is treated here as a statistical "reaction" between potential consumers and the *i-th* model, equivalent to the law of mass action. Hence, the present model is confined to an anonymous market. Note that Eq. (24) expresses Say´s theorem, which suggests that supply creates its own demand. This is the case here, because the more products $x_i$ are available the more will be purchased.

With this approximation the evolution of the density of available products $x_i(\tau)$ is governed by:

$$\frac{dx_i}{d\tau} = \eta_i \gamma_i x_i \psi$$
(25)

where we used Eq.(21).

## 2.3. The Spreading Process

In order to characterize the first purchase process, also known as diffusion process, we introduce the market penetration (adopter density):

$$n(t) = \frac{N_A(t)}{M}$$

(26)

where $N_A(t)$ is the cumulative number of adopters at time step $t$. The total sales due to first purchase is related to the change of the total number of adopters according to:

$$y_{tf}(t) = \frac{dn(t)}{dt}$$

(27)

Next to the first purchase process we have to take into account two other processes, related to the repurchase of the good. Due to the average finite lifetime $t_P$ of the models, consumers are forced to replace the good, unless it is not substituted by a complete new innovation. This process is denoted as replacement purchase, $y_{tR}(t)$. It is also possible to purchase more than one unit. This process is known as multiple purchase, with the total multiple purchase sales $y_{tm}(t)$. The total sales, also called the product life cycle, must therefore be given by:

$$y_t(t) = y_{tf}(t) + y_{tR}(t) + y_{tm}(t)$$

(28)

As will be shown below, the repurchase process of a good can be derived from the first purchase process $y_f(t)$. Therefore we start with a study of the adoption process.

### 2.3.1. Bass Diffusion

Suppose the introduction price $\mu_0$ of a durable good is such that $\mu_0 << \mu_m$. In this case the market is called homogeneous and the market volume is equal to the market potential. The product diffusion for a homogeneous market can be described by the well-known Bass model and is therefore denoted here as Bass diffusion. The simplest version is given by [5]:

$$y_{tf}^{B} = \frac{dn_B(t)}{dt} = \mathrm{A}\psi_f(t) + \mathrm{B}n_B(t)\psi_f(t)$$

(29)

where the time dependent density of potential consumers is determined by:

$$\psi_f(t) = 1 - n_B(t)$$

(30)

and $n_B(t)$ is the density of adopters due to the Bass adoption process. The first term of the differential equation describes a spontaneous purchase by potential adopters, where $A$ is the so-called innovation rate. The second term is due to the word-of-mouth effect. The density of adopters increases with an imitation rate $B$, proportional to the product of the densities of adopters and potential adopters.

For the case $\mu_0 >> \mu_m$, due to the income limitation, just those agents who can afford the product contribute to the number of potential adopters. Hence the density of potential adopters is limited by the market volume $v(\mu_0)$:

$$\psi_f(t) = v(\mu_0) - n_B(t)$$

(31)

Obviously, after introduction of a good the purchase process starts always with Bass diffusion, because $v(\mu_0) \neq 0$. Bass diffusion is therefore considered as the spreading process mentioned in the introduction, since it does not require any adaptation of the good. For a sufficiently long time Bass diffusion approaches its stationary state at $n_B = v(\mu_0)$. The adopter density evolution due to the Bass diffusion is given by:

$$n_B(t) = \frac{1 - e^{-(A+B)t}}{\left(1 + \frac{B}{A}e^{-(A+B)t}\right)^2} n_{B0}$$

(32)

while $n_{B0} = v(\mu_0)$. The total first purchase unit sales caused by Bass diffusion can be written as:

$$y_{tf}^{B}(t)=\frac{A(A+B)^2 e^{-(A+B)t}}{\left(A+Be^{-(A+B)t}\right)^2}n_{B0}$$
(33)

**2.3.2. The Repurchase Process**

We want to establish the repurchase processes for the case that the market is governed by Bass diffusion. In order to specify replacement and multiple purchase we take advantage from the first purchase process, $y_{tf}^{B}(t)$.

The key idea of replacement purchase is that the time to a replacement of the good can be described by a probability distribution, $\Gamma(t)$ of product failure over the population of units [12,13]. Replacement purchase is therefore determined by the chance of a product failure at *t''*, times the number of sales at *t-t''*. The integration over all possible lifetimes delivers:

$$y_{tR}^{B}(t)=R\int_{0}^{t} y_{tf}^{B}(t-t'')\Gamma(t'')dt''$$
(34)

where *R* is the fraction of previous sales suffered form replacement purchase. Note that Eq.(34) implies that replacement purchase is recurrent with $t_p$. This approach of replacement purchase is valid for the case that the impact of a reseller market can be neglected. Otherwise, the product owner could sell the old product and repurchase a new one at a time step not correlated with $t_p$.

The product failure probability $\Gamma(t'')$ can be approximated amongst others by a Gaussian distribution around the average lifetime, $t_p$. Confining our interest to the main contribution of replacement purchase, $\Gamma(t'')$ can be considered as a very sharp failure distribution around $t_p$. In this case the probability distribution can be approximated by a Dirac delta function, $\Gamma(t'')=\delta(t_p)$. For the first fundamental of the recurrent repurchase process, Eq. (34) turns for $t \geq t_p$ into:

$$y_{tR}^{B}(t)\cong Ry_{tf}^{B}(t-t_p)$$
(35)

else, $y_{tR}^{B}(t)=0$. Therefore, replacement purchase induces periodic variations of the unit sales with a periodicity given by the average product lifetime.

Any other repurchase decision, not correlated with the first purchase fundamental wave, is denoted here as multiple purchase. In difference to replacement purchase, multiple purchase must be proportional to the actual number of adopters, $n_B(t)$. Hence the sales can be approximated by:

$$y_{tm}^{B}(t)=Qn_B(t)$$
(36)

where *Q* is a multiple purchase rate.

**2.4. The Evolutionary Process**

In this chapter we want to focus on the period after Bass diffusion, while we consider the case that the mean product price $<\mu> >>\mu_m$.

iv) We assume that the price distribution $P_\mu$ is localized around the mean price.

The idea behind this assumption is that business units have the tendency to make their price decisions on the basis of the price of their competitors. This assumption is known in the economic literature as the "law of one price" (Jevons law). It implies that the $\eta_i$ are similar.

After a sufficiently long time, first purchase demand due to Bass diffusion will disappear, $\psi_f(t)\rightarrow 0$, while $n_B(t)\rightarrow v(<\mu>)$. In the phase after Bass diffusion, replacement demand will be small, since the lifetime $t_p$ of a durable good is long. Therefore we neglect periodic variations of the demand function and consider the repurchase process as governed by multiple purchase. Since multiple purchase is proportional to the number of adopters suffered from to the initial Bass diffusion process, $n_{B0}\sim v(<\mu>)$, repurchase demand must be proportional to the market volume. New potential consumers therefore occur in the period after Bass diffusion with a mean demand rate:

$$d(\langle\mu\rangle) = qv(\langle\mu\rangle)$$
(37)

while $q$ is a mean creation rate due to multiple purchase decisions. The parameter $q$ can be considered to be also a function of the price. However, for durable goods multiple demand is treated as a price independent constant, because the marginal utility for the simultaneous use of many models (>5) of the same durable good (TV's, PC's, VCR's etc.) will disappear. Hence $q$ can be approximated by a constant.

The total density of potential adopters on the short time scale is governed by the balance:

$$\frac{d\psi}{d\tau} = d - y_t$$
(38)

where the first term represents the increase of potential adopters due to repurchase decisions and the second the decrease due to the total purchase of the good. For the stationary state $d\psi/d\tau=0$ we obtain:

$$d = y_t$$
(39)

i.e. the total number of sold products is equal to the total repurchase demand. This relation implies that the mean demand $d(<\mu>)$, must be a constant on the short time scale according to Eq.(18).

The stationary density of potential adopters is given by:

$$\psi_S(\langle\mu\rangle) = \frac{q}{\sum\eta_i x_i} v(\langle\mu\rangle) = \psi_0 v(\langle\mu\rangle)$$
(40)

using Eq.(24). In order to test the stability of this state the density is perturbed by a small quantity $\delta\psi(\tau)$, such that $\psi(\tau)=\psi_S+\delta\,\psi(\tau)$, and $\delta\,\psi(\tau)\sim e^{\lambda\tau}$. Inserting this relation into Eq.(38) we get:

$$\frac{d\delta\psi(\tau)}{d\tau}=-\delta\psi\sum\eta_i x_i$$
(41)

Since $\sum\eta_i x_i>0$, we obtain that $\lambda<0$. Hence, small density fluctuations disappear and $\psi_S$ is a stable state for a given mean price.

Taking advantage from this result Eq.(25) can be written as:

$$\frac{d\eta_i\psi_S x_i(\tau)}{d\tau}=f_i\eta_i\psi_S x_i(\tau)$$
(42)

where both sides are multiplied with $\eta_i$ and $\psi_S$. From Eq.(24) follows that $\eta_i\psi_S x_i(\tau)=y_i(\tau)$, neglecting high frequency fluctuations proportional to $\delta\psi(\tau)$, since they disappear rapidly on the short time scale according to Eq. (41). The rate $f_i$ becomes:

$$f_i=\eta_i\gamma_i\psi_0 v(\mu_i)$$
(43)

where we used that the market volume of the *i-th* product is determined by $\mu_i$, and $\mu_i$ is close to $<\mu>$.The constraint Eq.(18) that $y_t$ is a constant on this time scale can be satisfied by adding a constant growth rate $\zeta$ to Eq. (42) such that:

$$\frac{dy_i(\tau)}{d\tau}=\left(f_i-\zeta\right)y_i(\tau)$$
(44)

From Eq.(18) we obtain:

$$\zeta=\langle f\rangle=\frac{\sum_i y_i f_i}{y_t}$$
(45)

Rewriting Eq.(42), the sales evolution of the *i-th* model is determined by the replicator equation:

$$\frac{dy_i(\tau)}{d\tau}=\left(f_i-\langle f\rangle\right)y_i(\tau)=r_i y_i(\tau)$$
(46)

where we have introduced the growth rate of the *i-th* business unit in terms of unit sales, $r_i$.

Before we proceed, we want to discuss this result. From the theory of evolution the parameter $f_i$ in the replicator equation is known as the fitness. Therefore, we want to denote $f_i$ here as the product fitness. It represents the effective reproduction rate of a model. The replicator equation indicates that competing models suffer from the evolutionary VSR-process.

When the spreading process (Bass diffusion) is over, the market is dominated by the repurchase process. Because in this phase the aggregate sales are constant on the short time scale, a selection pressure is created on the models. Since potential consumers have the choice between different models, the competition between them can be described by the replicator equation. It implies that the sales of those models which product fitness exceeds the mean fitness $\langle f \rangle$ are amplified. As a consequence, evolutionary competitive advantages lead to an advanced growth of the corresponding business units. The business units are therefore forced to increase the product fitness if they want to survive.

Note that the fitness function contains contributions from both sides of the market, demand and supply. The fitness is essentially determined by three parameters. It depends on the preference parameter $\eta$, the market volume as a function of the product price $\mu$, and the reproduction coefficient $\gamma$. They span a three-dimensional fitness space. The latter two can be varied within a short time period for example by marketing activities or output variations. Varying the product utility is related to innovations which requires more time, usually in the order of years. Although the adaptation process takes place simultaneously in the entire fitness space, we want to focus in this paper on the price and reproduction coefficient.

According to Eq.(20) and Eq.(39) the relation between aggregate supply and aggregate demand reads:

$$s_t = \langle \gamma' \rangle d$$

(47)

As discussed in Appendix A in the short term the mean reproduction parameter for profit maximizing business units is close to zero and thus $\langle \gamma' \rangle \approx 1$. Hence, aggregate supply is nearly equal to aggregate demand. The equivalence between demand and supply flow is called in standard microeconomics an equilibrium state [14]. The consumer market can therefore be considered to be in a quasi-equilibrium state on the short time scale.

In the long term, however, capacities and hence the output of the business units can be increased by investments. As shown in Appendix A, for a sufficiently long time interval, the mean reproduction coefficient is positive and proportional to the number of investments during that time interval. Because the reproduction coefficient and also the fitness is proportional to the profit per unit $g$, products with $g \leq 0$ will disappear from the market and the surviving products must have a $g > 0$, which is well-known economic fact.

The mutual competition between the products is an adaptation process known in the theory of evolution as evolutionary "arms race" [15]. Note that this "arms race" takes place on the demand as well as on the supply side. On the demand side a competitive advantage can be achieved by higher utility properties of the product or a better price. On the supply side the fitness advantage is due to investments increasing the output, for example due to an improved production technology.

In the next sections we want to derive a number of consequences of this approach.

#### 2.4.1. A Constant Fitness Advantage

We want to discuss the impact of a constant fitness advantage over a long time period. The market can be reduced to the two product case, $i=1,2$. The first product has fitness $f_1$, while all other models can be summarized to the second product with fitness $f_2$, such that $f_1 > f_2$. The key variable of our consideration is the market share of sold products:

$$m_i = \frac{y_i}{y_t}$$
(48)

Scaling the replicator equation Eq.(46) by $y_t$ we obtain for the evolution of the market shares:

$$\frac{dm_1}{d\tau} = \left(f_1 - f_2 m_2\right) m_1 - f_1 m_1^{\,2}$$
(49)

where we used that the mean fitness can be written according to Eq.(45) as:

$$\langle f \rangle = m_1 f_1 + m_2 f_2$$
(50)

With

$$m_2 = 1 - m_1$$
(51)

the evolution of the market share is governed by the Fisher-Pry equation [16]:

$$\frac{dm_1}{d\tau} = \theta m_1 (1 - m_1)$$
(52)

where the fitness advantage is

$$\theta = f_2 - f_1$$
(53)

The differential equation can be solved by a separation of the variables. Taking advantage from the assumption that $\theta$ is a constant, the time evolution of the market shares on the long time scale has the form:

$$\ln\left(\frac{m_1}{m_2}\right) = \theta \varepsilon t + C_m$$
(54)

while $C_m$ is an integration constant. This result suggests that the relation $m_1/m_2$ must be a linear function plotted in a half- logarithmic diagram.

We can conclude that the accumulation of a constant fitness advantage $\theta$, leads to a replacement of the current products over a long time period according to a logistic law. The fitness advantage may be for example due to a higher preference compared to the other products $\eta_i > \eta_j$, or a permanent excess supply while the price is comparable. It can be further concluded that if the fitness of a new product entering the market is strictly less than the mean fitness, the new product will disappear.

### 2.4.2. The Price Distribution

In this chapter we want to derive a price distribution $P_\mu$. It is determined by the relative abundance that a product is purchased in the price interval $\mu$ and $\mu+d\mu$, while the mean price can be given similar to Eq.(15) by:

$$\langle \mu \rangle = \int_0^\infty P_\mu(\mu')\mu' d\mu'$$

(55)

In order to model the price distribution we take advantage from assumption iv) and introduce the price difference:

$$\Delta\mu_i = \left(\mu_i - \langle \mu \rangle\right)$$

(56)

The fitness of the *i-th* product can be expanded as:

$$f(\mu_i) = f\left(\langle \mu \rangle\right) + \frac{df\left(\langle \mu \rangle\right)}{d\mu}\Delta\mu_i$$

(57)

Inserting this result in Eq.(45) we obtain that the mean fitness averaged over all supplied products can be given by the fitness at mean price, since $<\Delta\mu>=0$:

$$\langle f \rangle = f\left(\langle \mu \rangle\right)$$

(58)

Eq. (56) suggests that the price evolution of the products can be separated into the evolution of price deviations and of the mean price:

$$\frac{d\mu_i}{d\tau} = \frac{d\Delta\mu_i}{d\tau} + \frac{d\langle \mu \rangle}{d\tau}$$

(59)

In order to determine the price distribution $P_\mu$, let us focus first on price deviations around the mean price. The evolution of the mean price is considered below.

Based on assumption iii) we assume for large sales variations:

v) When product sales decrease, the business units respond by decreasing the product price with respect to the mean price and vice versa.

The business units are forced to adopt the price on large sales fluctuations because they have a limited capacity range (Appendix A).The rule reads:

$$\frac{d\Delta\mu_i}{d\tau} \sim sign\left(\frac{dy_i}{d\tau}\right)$$
(60)

Note that we assumed no absolute relationship between sales and price variations. Eq.(60) still holds, if we scale within the sign-function by the positive variable $y_i$. Because the sales are governed by the replicator equation and we can write:

$$\frac{d\Delta\mu_i}{d\tau} \sim sign\left(\frac{1}{y_i}\frac{dy_i}{d\tau}\right) \sim sign\left(f(\mu_i) - \langle f \rangle\right) = sign(r_i)$$
(61)

The present theory therefore suggests that price variations are caused by fitness variations of the products around the mean fitness. Taking advantage from Eq.(57) the relation turns into:

$$\frac{d\Delta\mu_i}{d\tau} \sim -sign\left(\left|\frac{df}{d\mu}\right|\Delta\mu_i\right)$$
(62)

where we used that the first derivative of the market volume with respect to the price is always negative:

$$\frac{df(\mu)}{d\mu} \sim \frac{dv(\mu)}{d\mu} < 0$$
(63)

However, we have not specified the magnitude by which manufacturers respond on supply variations. This can be done statistically. We consider the price variation as consisting of a deterministic part and a stochastic contribution of the form:

$$\frac{d\Delta\mu}{d\tau} = F(\Delta\mu) + \delta\mu(t)$$
(64)

where the deterministic part is determined by the response on sales fluctuations Eq.(60):

$$F(\Delta\mu) = -bsign(\Delta\mu)$$
(65)

The proportionality constant $b=<\Delta\mu>_\tau/\Delta\tau$ can be determined by a time average over the response of the business units within a given time period. The random price contribution $\delta\mu(\tau)$ is treated as white noise with mean value and time correlation:

$$\langle \delta\mu(\tau) \rangle_\tau = 0$$
$$\langle \delta\mu(\tau), \delta\mu(\tau') \rangle_\tau = D\delta(\tau - \tau')$$

(66)

The brackets with index $\tau$ indicate the time average and $D$ is a noise amplitude.

The price evolution given by Eq.(64) is a stochastic Langevin equation. The first term can be interpreted as a restoring force $F(\Delta\mu)$, which keeps the price of the models close to the mean price. Based on the corresponding Fokker-Planck equation, the stationary price distribution can be given by (Appendix B):

$$P_\mu(\Delta\mu) \sim \exp\left( -\frac{2b}{D}|\Delta\mu| \right)$$

(67)

The price distribution in this model is therefore a Laplace (double exponential) distribution. In a semi-log plot, the Laplace distribution has a tent shape around the mean price, $<\mu>$.

Because the business units do not respond on arbitrary small sales variations with price adaptations, small price variations are not directly related to sales variations. Hence the Laplace distribution describes only large price variations around the mean price. Such a distribution is known as fat tails [17].

### 2.4.3. The Size and Growth Rate Distribution of the Business Units

In this section we want to focus on the size and growth rate distribution of the business units, and define the size of a business unit by its unit sales. The size distribution of the business units $P_y$, is determined by the probability to find the unit sales of a business unit $y_i$ in the interval $y$ and $y+dy$.

Scaling Eq.(46) by $y_t$, the time evolution of the relative abundance of the unit sales is governed on the short time scale by:

$$\frac{dP_y(y)}{d\tau} = r\, P_y(y)$$

(68)

From Eq. (18) and (46) follows that the mean growth rate on the short time scale is:

$$\frac{1}{y_t}\frac{dy_t}{d\tau} = \frac{1}{y_t}\sum_i r_i y_i = \langle r \rangle = 0$$

(69)

Although the mean growth rate is zero, the growth rates of the individual business units fluctuate around the mean value, governed by a growth rate distribution $P_r$. Eq.(68) indicates, that the size evolution of the business units is determined by a multiplicative stochastic process. The central limit theorem suggests in this case the size distribution of the business units is given for a sufficiently long time by a lognormal probability distribution function (pdf) of the form:

$$P_y(y,t) = \frac{1}{\sqrt{2\pi t}\,\omega y} \exp\left( -\frac{\left(\ln(y/y_0) - ut\right)^2}{2\omega^2 t} \right)$$

(70)

where $u$ and $\omega$ are free parameters and $y/y_0$ is the size of the business unit scaled by the size at $t=0$. Note that the size distribution broadens on the long time scale, but can be considered as stable on the short time scale.

The growth rate distribution $P_r(r)$, determines the relative abundance of a business unit to have a growth rate in the interval $r$ and $r+dr$. Note that fluctuations of the price are directly related to variations of the grow rate Eq.(61). The growth rate distribution of the unit sales can therefore be obtained directly from the price distribution by changing variables. We obtain (Appendix C):

$$P_r(r) \sim \exp\left(-|r|\right)$$

(71)

where

$$r \cong \log\left( \frac{y(\tau+1)}{y(\tau)} \right)$$

(72)

is determined by the logarithm of the unit sales. Hence, the evolutionary model suggests that the stationary growth rate distribution is given also by a Laplace (double exponential) distribution.

### 2.4.4. The Mean Price Evolution

As suggested in Eq. (59), also the mean price evolves in this evolutionary model. Because the relative abundance of the sales as a function of the price is governed by the replicator equation, we can write for the evolution of the price distribution:

$$\frac{dP_\mu}{d\tau} = \left(f(\mu) - \langle f \rangle\right) P_\mu$$

(73)

With Eq.(55) we obtain for the evolution of the mean price

$$\frac{d\langle \mu \rangle}{d\tau} = \int_0^\infty \frac{dP_\mu(\mu')}{d\tau} \mu' d\mu'$$

(74)

and using Eq.(73):

$$\frac{d\langle \mu \rangle}{d\tau} = \int_0^\infty P_\mu(\mu')\mu'\big(f(\mu') - \langle f \rangle\big)d\mu'$$
(75)

This relation becomes:

$$\frac{d\langle \mu \rangle}{d\tau} = \int_0^\infty \mu' f(\mu') P_\mu(\mu') d\mu' - \langle \mu \rangle \langle f \rangle$$
(76)

Applying Eq.(57) the mean price is governed by:

$$\frac{d\langle \mu \rangle}{d\tau} = \frac{df(\langle \mu \rangle)}{d\mu} Var(P_\mu)$$
(77)

where the price variance is defined as:

$$Var(P_\mu) = \int P_\mu(\mu')\mu'^2 \, d\mu' - \left(\int P_\mu(\mu')\mu' d\mu'\right)^2$$
(78)

Eq.(77) is a well-known result in the evolutionary theory [18,19].

The stationary solution of Eq. (77) is determined here by

$$\frac{d\langle \mu \rangle}{d\tau} \sim \frac{df(\langle \mu \rangle)}{d\mu} \sim \frac{dv(\langle \mu \rangle)}{d\mu} = 0$$
(79)

because the variance is always positive. From the market volume Eq.(9) follows that the stationary price is given by $\mu_m$. Because the market volume has a maximum there, $\mu_m$ is globally stable. For a stable price distribution we obtain that the mean price approaches $\mu_m$ according to (Appendix D):

$$\langle \mu(t) \rangle = \mu_0 e^{-at} + \mu_m$$
(80)

where $\mu_0$ is the price at $t=0$. The parameter $a$ is denoted here as the price decline rate and is treated as a constant on the long time scale. The evolutionary theory therefore suggests that the average price for durable goods decays exponentially and approaches $\mu_m$ asymptotically on the long time scale.

### 2.4.4. Gompertz Diffusion

As consequence of the price decline is that the market volume and hence the number of adopters increase with time. The increase of the adopter density can be interpreted as a diffusion process. For a localized price distribution, the increase of the adopter density is equal to the increase of the market volume at mean price:

$$\frac{dn_G(t)}{dt} = \frac{dv\left(\langle \mu(t) \rangle\right)}{dt}$$
(81)

and hence:

$$n_G(t) = v\left(\langle \mu(t) \rangle\right) - v(\mu_0)$$
(82)

where the integration constant is the market volume suffered from Bass diffusion. Applying Eq.(9) the evolutionary approach implies that the adopter density evolves as:

$$n_G(t) = n_{G0} \exp\left( - \frac{\left(\langle \mu(t) \rangle - \mu_m\right)^2}{2\Theta^2} \right)$$
(83)

while

$$n_{G0} + n_{B0} = 1$$
(84)

With Eq.(82), the evolutionary model suggests that the diffusion process is determined by Gompertz equation:

$$n_G(t) = n_{G0} \exp\left(-ke^{-2at}\right)$$

(85)

with

$$k = \left( \frac{\mu_0}{2\Theta^2} \right)^2$$
(86)

Hence, the evolutionary diffusion process corresponds to Gompertz diffusion. Gompertz adoption rate reads:

$$y_{tf}^G(t) = \frac{dn_G(t)}{dt} = 2akn_G(t) \exp\left(-2at\right)$$
(87)

### 2.5. The Product Life Cycle of a Durable Good

Both diffusion processes can be considered as independent. Bass diffusion is interpreted as a spreading process and Gompertz diffusion as due to an evolutionary price adaptation process. Hence, the total adopter density is determined by:

$$n(t,t') = n_B(t) + n_G(t')$$
(88)

where we assumed that the spreading process starts at introduction $t_0=0$. The evolutionary process is shifted by $\Delta t_0$, such that:

$$t' = t + \Delta t_0$$
(89)

Now we are able to establish the product life cycle. Based on Eq.(28) the contribution to the product life cycle due to Bass diffusion is:

$$y_t^B(t) = y_{tf}^B(t) + Qn_B(t) + Ry_{tf}^B(t - t_p)$$
(90)

where the density and adoption rate are given by Eq. (32) and Eq.(33). Hence Bass diffusion requires three free parameters $n_{B0}=\nu(\mu_0)$, $A$ and $B$. And the product life cycle needs the additional parameters $Q$, $R$ and $t_p$.

Equivalently the contribution to the product life cycle caused by Gompertz diffusion is:

$$y_t^G(t') = y_{tf}^G(t') + Q'n_G(t') + R'y_{tf}^G(t' - t_p')$$
(91)

Gompertz diffusion consists also of three free parameters $n_{G0}$, $k$ and $a$, supplemented by the free parameters $Q'$ and $R'$ for multiple and replacement purchase. The product lifetime can in general be different for both adoption processes, $t_p \neq t_p'$. The aggregate unit sales become:

$$y_t(t,t') = y_t^B(t) + y_t^G(t')$$
(92)

Thus, the product life cycle can be expected to exhibit two characteristic periodic waves due to the two diffusion processes.

## 3. Comparison with Empirical Results

The aim of the present model is not to make precise statistical forecasts of the evolution of a good. Instead we want to show that a evolutionary theory allows a qualitative explanation of a number of empirical results within a single theoretical framework. For this purpose we want to compare the model with empirical results. The evolutionary model made the following predictions that can be compared with empirical data:

I) The size distribution is lognormal and the growth rates exhibit a Laplace distribution.

II) The price of the business units are governed by a Laplace distribution for large price deviations (fat tails).

III) The diffusion process consists of a Bass and a Gompertz regime. Bass diffusion is not directly related to the price. Gompertz diffusion, however, has a tight relationship to the mean price evolution, which is predicted to show an exponential decline.

IV) The aggregate unit sales can be separated into a Bass and a Gompertz contribution. The product life cycle can be expected to approach a stationary state with periodic variations due to a replacement purchase of the good.

V) A product with a constant fitness advantage replaces competitive models, such that the market share of the unit sales is governed by a logistic law.

### 3.1. The Size, Growth Rate and Price Distribution

Empirical investigations of the growth rate distribution do not consider business units, but firms. Since the number of business units within a firm may be quite large, the comparison with empirical data is of qualitative nature. Statistical studies on the growth rate distribution of firms showed for manufacturing firms, that the sales pdf is located around $r=0$ as suggested by the model. Note that unit sales and sales in financial terms can be used equivalently, since the price can be represented by the mean price. It was found empirically that the growth rate distribution can be described by a Laplace distribution [20]. This result is in agreement with the growth rate distribution derived in this theory.

A detailed analysis revealed that the growth rate distribution depends on the firm size and can be approximated by a Subbotin distribution [21,22]. It was found that the standard deviation of the growth rate distribution scales as $y^{-\beta}$, with $\beta \approx 0.2$. Since a derivation of the size dependence of the growth rate distribution is beyond the purpose of the paper a direct comparison with this relation is not possible. One explanation for this relationship is that the size dependence is due to a self-similar structure of the business units within firms. The growth of larger firms fluctuate therefore less than smaller ones [23]. This idea fits into the evolutionary view of this paper, since firms build up out of diverse business units have an evolutionary competitive advantage.

The model also suggests an intimate relationship between firms and their products. The growth rate distribution of the products must be equivalent to the corresponding distribution of the business units. The empirical investigation of products in the worldwide pharmaceutical industry revealed that the growth rate distribution of the products is indeed nearly equivalent to the growth rate distribution of the firms, which confirms the present model [24].

That the firm size distribution (in sales) should be lognormal is known as the law of proportionate growth (Gibrat's law [25]). The present model, however, suggests that the lognormal distribution is strictly valid only for products (business units) but not for firms. This result was confirmed empirically by Growiec et al. [26]. They found a lognormal distribution for pharmaceutical products, but for firms the lognormal distribution has a power law departure in the upper tail, because large firms consist of a number of business units (products).

The statistical determination of the price distribution requires a large amount of data. They can be obtained from trade actions at the stock exchange, which allows a detailed study of the price distribution, known for example for shares, bonds, currencies etc. Many of these price distributions were found to show fat tails [17]. Similar results were found for non-durables as for example for electricity prices [27]. However, there is no a clear empirical evidence of a fat tailed price distribution for durable goods.

### 3.2. The Diffusion Process and the Product Life Cycle

While the diffusion process can be characterized by the market penetration, the product life cycle is determined by the total unit sales. Usually empirical investigations focus on aggregate data. Unfortunately aggregate data contain all processes including Bass and Gompertz diffusion. The product life cycle require at most 15 free parameters. Therefore the characterization of the diffusion process based on aggregate data is quite complex. The empirical investigations discussed here are arranged such, that we start with those examples, having the lowest number of required free parameters.

Both Bass and Gompertz diffusion are used in the literature as tools to forecast the market penetration. The evolutionary theory suggests that for durable goods they can be expected to be evident both. However, it is often difficult to separate the empirical data into the two diffusion regimes. Bass diffusion can be expected to dominate durable goods with mostly industrial applications, since firms are usually not limited by the price. On the other hand, Gompertz diffusion should be evident in the diffusion process of expensive consumer goods. In order to separate between the two diffusion regimes we take advantage from the predicted intimate relationship between price and market penetration of the good.

The mean price evolution is determined by three unknown parameters and is given by Eq.(80). In order to simplify the fit procedure, we introduce the following price function:

$$\mu'(t') = \frac{\mu(t') - \mu_m}{\mu_0} \cong \frac{p(t') - p_m}{p_0} \sim e^{-at'}$$

(93)

while $t'$ is given by Eq.(89). It starts when an exponential decrease of the price evolution is evident. The case $\Delta t=0$ indicates in this paper, that the start of the price decline cannot clearly be given.

The price function Eq.(93) has two advantages:

1. For short time periods, the mean income scales out, because it changes slowly. Therefore we can apply the nominal market price $p(t')$ instead of the real price $\mu(t')$.
2. According to Eq.(80) the scaled price function $\mu'(t')$, should be a linear function in a semi logarithmic plot. Hence, with an appropriate choice of the parameter $p_m/p_0$, the empirical data arrange into a linear function, while its slope determines the price decline rate $a$.

In order to compare empirical data with the presented model, the fit procedure starts by specifying $\Delta t$ and the corresponding introduction price $p_0$ in Eq.(93). Applying a least square fit for the two other unknown parameters $a$ and $p_m/p_0$, we obtain the solid lines for $\mu'(t')$ displayed in the figures. While for short time periods scaling by the mean income in Eq.(92) can be neglected, for long time periods the mean income evolution has to be taken into account, $I=I(t')$. The mean price can be evaluated from $\mu(t')=p(t')/I(t')$, where $p(t')$ is the empirical nominal price. For the annual average income we have taken advantage from the empirical data given by Silva and Yakovenko [8]. The annual income can be approximated by the function:

$$I(t') = I_0(1+T)^{t'}$$

(94)

where $t'$ is given in years and $T=0.05$, $I_0=4400$ US\$ as displayed in Fig.2. This relation suggests an average increase of the annual US-income of about 5%. This procedure was applied only to Colour TV sets.

For the diffusion process the parameter $a$ is considered as fixed for Gompertz diffusion. Since the parameter $k$ shifts the diffusion curve in time, it can be obtained from a least square fit by varying $n_{G0}$. Empirical studies determine the market penetration as a function of the number of households $N_H$ in a country. Therefore the market penetration is mostly displayed as percentage of households. The maximum market penetration in terms of households is expressed here as: $n_{max}=M/N_H$. In difference to the theoretical derivation the maximum market penetrations is therefore given by $n_{max}=n_{B0}+n_{G0}$.

In market penetration data the impact of Bass diffusion is sometimes very small. Its contribution to the sales in the product life cycle is, though, evident. Therefore the parameters $A$, $B$ and $n_{B0}$ are determined from the empirical unit sales. The parameter $n_{B0}$ was adjusted such that iteratively the market penetration and first purchase units sales are in good coincidence.

The other parameters associated with repurchase and multiple purchase come only into play for a consideration of the unit sales over a long time horizon. Since explicit data from the repurchase processes are not available for the presented examples, we have estimated these parameters.

**Colour TV sets**

As a first example we want to consider the diffusion of Colour TV sets in the USA. This example has been chosen, because empirical data are known for the market price, market penetration and first purchase sales over a long time period. Here a minimum number of free parameters are required. The price dependence can be fitted by $a$ and $p_m/p_0$. The market penetration and first purchase sales by five parameters, $n_{B0}$ ,$A$, $B$ for Bass diffusion and $n_{G0}$ and $k$ for Gompertz diffusion. For the Colour TV sets, the scaled price $\mu'(t')$ and the market penetration $n(t')$ were fitted to the empirical data given by Vanston [28,29]. They are displayed in Fig.3, while the free parameters are summarized in Table1. Because the time period was more than thirty years, we have taken advantage from Eq.(94).

Displayed in Fig.4 are the empirical data of the first purchase sales of Colour TV-sets [28]. Since first purchase sales are determined by the first derivative of the market penetration Eq.(27), this function is completely determined by the five parameters of the market penetration. The small peak around 1958 in the sales is caused by Bass diffusion while the second is due to Gompertz diffusion. The large sales variation around 1972 is probably due to a price fluctuation during the same time period evident in Fig.(3). This example exhibits the predicted intimate relationship between price and market penetration, since the exponential price decline as displayed in Fig.3 governs the market penetration and first purchase sales.

**Fax Machine**

A textbook example is the diffusion of FAX machines in the USA [14]. Economides and Himmelberg [30] investigated the diffusion process in detail. This example is interesting in two aspects. First, the exponential price decline starts a few years after introduction of the product as displayed in Fig. 5. Also shown in this figure are the empirical data of the market penetration. Obviously Bass diffusion dominated the first years after introduction of the product, followed by Gompertz diffusion, shifted by $\Delta t \approx 4$ years. The solid lines represent a fit to the empirical data with the parameters summarized in Table 1. The price evolution during Bass diffusion is not derived in this model and therefore not further considered.

Second, the authors also investigated the growth rate of the installed base shown in Fig.6. These data are more instructive than the aggregate unit sales, since they represent first and multiple purchase, but not replacement purchase. The solid line in Fig.6 is therefore determined by Eq.(92) with $R=R'=0$.

Remarkable is that Bass diffusion exhibits a high multiple purchase parameter, while for Gompertz diffusion we estimate, $Q' \approx 0$. This result can be interpreted as follows: A FAX machine is a network product first applied by the industry. For a standard consumer, it is not useful to own more than one FAX machine. Industrial users, however, repurchase this product in order build their own local networks. Because firms are usually not limited by the price, Bass diffusion dominated the first three years followed by a nearly constant multiple purchase of the product, as suggested by the theory. Gompertz diffusion is related to the sales peak around 1988 in Fig. 6. There is no additional multiple purchase by the consumers beyond this peak. Because non-industrial users purchase only one FAX, the sales return to the level of multiple purchase by the industrial agents after Gompertz diffusion.

**Black & White TV sets**

Empirical data for the market penetration of Black & White (B&W) TV sets in the USA and the evolution of the real mean price was given by Wang [31]. Displayed in Fig.7 are the price function $\mu'(t)$ and the market penetration, where the solid lines are fits with the parameters given in Table 1.

This example was chosen, because the product life cycle is known for a long time period of about thirty years as displayed in Fig.8 [32]. Remarkable is that the sales exhibit periodic variations as suggested by the present model. Also displayed in Fig.8 is a fit of the sales according to Eq.(92), where the whole parameter space was applied (Table1).

According to the present model the product life cycle should consist of two periodic waves, associated with Bass and Gompertz diffusion. Although the empirical data could not fitted exactly, the periodic price variations are evident. The wave associated to Bass diffusion has peaks at 1950, 1959, 1968 and probably 1977. Hence, the product lifetime is about, $t_p \approx 9$ years. The other sales peaks at 1955, 1965 and 1973 are related according to the present theory to Gompertz diffusion with a product lifetime, $t_p' \approx 10$ years. Although the coincidence of the sales curves is not perfect, the results confirm the presented theory qualitatively.

Note that the sales of B&W TV's are interfered by the introduction of Colour TV's, which has its growth period ~ 1964-70 (Fig.3). The preference parameter of Colour TV's is certainly much higher than of B&W TV's, $\eta_{Colour} >> \eta_{B\&W}$. However, there is no rapid decrease of the sales evident in the B&W TV's sales data after introduction of Colour TV's. This result indicates that a market can separate into independent market segments if there is a considerable difference in the utility properties $\eta_i >> \eta_j$ and price $\mu_i >> \mu_j$. Note, that the mean price must be related to the corresponding market segment. The segments merge into each other when the mean price becomes comparable. This is the case after 1978, associated with a replacement of the B&W TV's by Colour TV's, not displayed here [31].

**Air Conditioners and Dryers**

Unfortunately, most empirical investigations focus either on the market penetration or on unit sales, neglecting the price evolution. While aggregate market penetration data can often be fitted by applying either Bass or Gompertz diffusion, the study of the product life cycle turns out to be the more complex, in particular when Bass and Gompertz diffusion are equally dominant. This seems to be the case for the diffusion of air conditioners and clothes dryers in the USA. The empirical data of the product life cycle of these goods were given by Steffens [32] and are displayed in Fig. 9 and 10. Also displayed is a fit of the unit sales according to Eq.(92) where we have neglected for simplicity the replacement contribution and set $\Delta t=0$. The product life cycle can be understood in view of the present evolutionary theory as a sequence of Bass and Gompertz diffusion. In order to indicate its presence, the first purchase sales from Bass and Gompertz diffusion are displayed in these figures by grey lines.

## 3.3. A Constant Fitness Advantage

The evolutionary model predicts a logistic growth of the sales market share of products with a constant competitive advantage. There are several examples, where logistic growth of market shares was found in economic data [33, 34, 35, 36]. Here, we want to confine to an example that allows the application of both parts of the evolutionary model; the characterization of the evolution of the entire market and the dominance of a product (here a standard): the VCR case.

**VCR**

Video Cassette Recorders (VCR) where introduced in the USA 1976 with the first standalone Betamax machine by Sony. A year later the VHS format entered the market. The evolution of these two dominant standards was studied in detail for example by Park [37]. Both formats were produced by a number of manufacturers. The price evolution of the two formats is displayed in Fig. 11, together with the mean price. The bars indicate the price range of the two VCR- standards [37]. The solid line represents a fit of Eq.(80) to the nominal prices. Obviously the mean price decreases in time, which indicates the action of the VSR-mechanism.

While the market penetration is not displayed here, the VCR unit sales are shown in Fig.12. Both Gompertz and Bass diffusion processes contribute successively to the total unit sales $y(t)$ (solid line) while we have taken advantage from the price decline in Fig.11. However in difference to previous samples (B&W TV's) a pronounced peak due to Bass diffusion is absent. Also displayed are the sales of the two formats, $y_{VHS}(t)$ and $y_{Beta}(t)$.

We treat the two standards as in the two-product case. The VHS-market share is given by:

$$m_{VHS}(t) = \frac{y_{VHS}(t)}{y_t(t)}$$

(95)

and $m_{Beta}(t)=1-m_{VHS}(t)$. Plotting the VHS market share in a half logarithmic diagram the theory suggests a linear function of the form:

$$\ln\left(\frac{m_{VHS}(t)}{m_{Beta}(t)}\right) = \theta t + C_m$$

(96)

This plot is performed in Fig.13. The solid line is a fit of Eq.(96) with $\theta \cong 0.22$ and $C_m=0$. The present model therefore suggests that the VHS format has a relative constant fitness advantage compared to the Betamax format. Using the total sales from Eq.(92) and applying the logistic equation Eq.(96) with $\theta=0.22$, the sales of the formats can be re-evaluated and are displayed by the dotted lines in Fig.12. The coincidence indicates that the present theory describes the relation between the sales of the formats and the total market qualitatively correct. Note that the fitness advantage is not due to a lower price, since the price range in Fig. 11 is comparable for both formats. Since $\theta$ is nearly constant, the success of the VHS format is most probably due to a preference advantage or a higher reproduction coefficient. The latter is more probable, because the number of VHS manufacturers considerably increases the number of Betamax producers [37].

In economic literature the source of the competitive advantage is assumed to be due to the network effect [38, 39]. This effect suggests that adopters have a utility advantage form a network product, which increases with the number of adopters [40]. Note that the network effect implies an increase of $\theta$ with time, due to the increased number of adopters of a format. Indeed at the end of the eighties an increase of $\theta$ is evident. It is responsible for the deviation of the predicted sales functions from the empirical data in Fig. 12 after 1986. An unambiguous attribution of this result to the network effect require however a specification of the network effect, not performed here.

## 4. Conclusion

In marketing research increasingly complex models based on Bass diffusion were developed in order to explain the product life cycle with economic decision variables [41,42,43]. The presented model offers a completely different approach. It suggests that the diffusion process of a durable good consists of a spreading process governed by Bass diffusion, and an evolutionary Gompertz diffusion regime associated with the expansion of the market volume with a decreasing price. In the latter diffusion phase the market is on the short time scale in a quasi-equilibrium state, associated with a nearly constant total output. Gompertz diffusion is the result of the competition between manufacturers (business units) for new potential adopters.

Gompertz diffusion for durable goods can be visualised as schematically displayed in Fig.(14). The fat line indicates the market volume, $v(\mu)$. It represents the effective demand curve of a good. The price distribution $P_\mu$ is governed by random price fluctuations. The present theory suggests that the price distribution can be described by a Laplace distribution for large price deviations (fat tails) with a mean price at $<\mu(t)>$. The mean price evolution is determined by a gradient dynamics. The evolutionary model suggests that the velocity by which the mean price decreases is proportional to the gradient of the fitness (market volume). The mean price approaches $\mu_m$ asymptotically governed by an exponential law (see insert, $\mu(t)$). Associated with this price decrease is an expansion of the market volume, respectively adopter density. The corresponding diffusion process is governed by Gompertz equation (see insert, $n(t)$). Displayed in the other insert is the product life cycle, characterized by the total unit sales $y_t(t)$ (fat line). It consists of first purchase $y_{tf}(t)$, multiple purchase $y_{tm}(t)$ and replacement purchase $y_{tR}(t)$ (dotted lines).

Due to the large parameter space, a general discussion of aggregate empirical results is complex without additional information about first- and repurchase processes. However, the model predicts a tight relationship between the exponential mean price decrease and Gompertz diffusion. This relation can be tested rigorously. The comparison with empirical data of the diffusion of B&W TV sets, Colour TV sets, Fax machines and VCR in the USA confirms this relationship.

More complex is the characterization of the long-term evolution of the sales. Because first purchase demand splits into a Bass and a Gompertz contribution the adoption process may create two sales peaks. Additionally, replacement purchase induces periodic variations of these peaks. The investigation of the empirical sales data of B&W TV sets verifies this prediction. Note that the finite lifetime of a durable good is usually of the order of 7-11 years. Periodic variations of this order are known in the economic theory as Juglar cycles. The present model is in agreement with the standard explanation of Juglar cycles as the result of investments associated with a replacement of products [44].

The model predicts that the price distribution is related a Laplace distribution caused by price decisions of competitive firms on sales variations. We want to emphasize the fundamental nature of competition in this theory. The mean price evolution is directly proportional to the variance of the price distribution (Eq.75). If the variance is zero, as in the case of a monopoly, the market is "frozen" into a state far away from a maximum fitness. That means, a market cannot relax into the state with maximum fitness, for the case of a monopoly. The negative impact of a monopoly on a free market is a well-known economic fact.

Note that previous evolutionary economic literature has its focus on the supply side of a market, in particular on the technological evolution of the business units ([45], [46]). The present approach takes the evolution on the demand side into account. The model explains the huge amount of products in a free market as the result of a permanent variation of models by the business units. Depending on their product fitness they can raise or fall. Products extinct and are replaced by a selection process caused by the consumers which can be described by a logistic law for a constant fitness advantage. The competition between the two VCR standards VHS and Betamax shows the predicted logistic replacement relationship. Similar to the evolution of species the theory also suggests that an evolutionary tree can be drawn for any man made good, from its first invention until its present applications.

Finally we want to mention the debate between an evolutionary and a neo-classic view of a market ([47]). The present approach suggests that both views merge into each other. It is a question of the time scale whether an equilibrium view or an evolutionary view is more appropriate. The situation is similar to the description of the continental drift. In every day life the earth crust is in quasi-equilibrium. Only on a long time scale the drift of the continents is evident. Equivalently a market can be treated on a short time scale as in equilibrium. On a long time scale we can see the evolution at work.

## Appendix A

## The Reproduction Parameter

This chapter is devoted to a consideration of the reproduction parameter $\gamma$. Let us consider an arbitrary business unit. The goal on the short time scale is to maximize the profit flow given by:

$$G(y) = py - K(s)$$
(A1)

where the total costs as a function of the supplied products $K(s)$ can be expanded as a function of the extra output $\gamma$ as:

$$K((1+\gamma)y) = K(y) + \left.\frac{dK}{dy}\right|_y \gamma y$$
(A2)

For $\gamma > 0$, the profit flow can be maximized by decreasing $\gamma \rightarrow 0$, since additional costs due to a mismatch between produced and sold products disappear. The extra supply must be altered until $s=y$ and thus $\gamma=0$. This can be done by a decrease of capacities. But because this variation is limited, on a large sales variation additionally the price is decreased. The case $\gamma<0$ suggests that a business unit sells more products than can be supplied. This effect decreases the number of available products $x$, which leads by means of Eq. (24) to a decrease of the sales in time. However the business units are not interested in decreasing sales, because it decreases the profit. Therefore, this situation is usually compensated by increasing the capacities and if this does not help by increasing the price. Hence they increase $\gamma$ such that $\gamma \rightarrow 0$. Therefore the reproduction coefficient will fluctuate in time around $\gamma(\tau) \approx 0$, in order to maximize the profit with an appropriate (limited) variation of the capacities.

The behaviour of the business units is used to derive the price distribution, and can be interpreted concerning the reproduction parameter as a restoring force $F_S(\gamma)$ keeping $\gamma$ close to zero. The dynamics of the reproduction coefficient can be approximated by:

$$\frac{d\gamma}{d\tau} = -F_S(\gamma)$$
(A3)

Because the restoring force must be zero at $\gamma=0$, we can expand $F_S(\gamma)$ and obtain up to the first order in $\gamma$:

$$\frac{d\gamma}{d\tau} \cong -\chi\gamma + \xi$$
(A4)

where $\chi$ is a mean compensation rate of the business units and $\xi$ indicates random fluctuations.

However, we have to take into account that the profit made by the products can be used to increase the capacities, known as investments. Investments are considered here as large positive jumps of the output within a short time interval $\Delta\tau$. They will lead also to sales fluctuations associated with price adaptations. Applying Eq.(19) the supply change of a single investment can be approximated by:

$$\Delta s = s(\tau + \Delta\tau) - s(\tau) \cong \gamma' y(\tau) - y(\tau) = \gamma y(\tau)$$
(A5)

We used that the output is equal to sales $s(\tau)=y(\tau)$ before the investment for a profit maximizing business unit and that the sales cannot immediately respond on the output jump. Thus the output increase leads also to a positive jump in $\gamma$.

For a number of investments within a period $\Delta\tau$ the output growth of a business unit can be written as:

$$\Delta s = \sum_{m}^{N_I(\Delta\tau)} \kappa_m$$
(A6)

where $\kappa_m$ indicates the impact of the *m-th* investment on the increase of the output and $N_I(\Delta t)$ is the number of investments during $\Delta\tau$. The output growths due to the accumulation of small jump like steps, while $\kappa_m$ determines the change of the capacity by the *m-th* investment. Approximating $\kappa_m$ by its mean value $\kappa$, we can write:

$$\sum_{m}^{N_I(\Delta\tau)} \kappa_m \approx N_I(\Delta\tau)\kappa$$
(A7)

For the evaluation of the number of investments we take advantage from the fact that investments have to be paid from the profit flow of the product. The time to repay the investment is known as the amortization time. For a mean amortization time $t_A$ the number of investments during $\Delta\tau$ can be approximated by:

$$N_I(\Delta\tau) \cong \frac{\Delta\tau}{t_A}$$
(A8)

Averaging $\gamma(\tau)$ over a time period $\Delta\tau$, delivers different results depending on the considered time interval. Choosing a very short time interval $\Delta\tau << \chi^{-1}$, the time averaged reproduction parameter $<\gamma>_\tau \neq 0$, because the business units are not relaxed from fluctuations. For $\chi^{-1} << \Delta\tau << t_A$ the impact of capacity changing investments can be neglected, while the business units have enough time to relax from fluctuations. Hence:

$$\langle \gamma \rangle_\tau \cong 0$$
(A9)

An average over this time period corresponds to a consideration on the short time scale. Choosing a very long time period $t_A << \Delta\tau$ a large number of investments are carried out leading to a considerable increase of the output and therefore:

$$\langle \gamma \rangle_t > 0$$

(A10)

while the index $t$ indicates a time average on the long time scale.

From Eq.(A5) and (A6) we obtain a relation for the mean reproduction parameter on the long time scale:

$$\langle \gamma \rangle_t \sim \frac{N_I}{y}$$
(A11)

The reproduction parameter can be further specified by the fact that the amortization time is related to the profit flow $G$. The higher the profit flow that can be made by the product the shorter the amortization time:

$$t_A \sim G^{-1}$$
(A12)

Writing the profit per unit as:

$$g = \frac{G}{y}$$
(A13)

we obtain that the average reproduction parameter is proportional to profit per unit:

$$\langle \gamma \rangle_t \sim g$$
(A14)

where we used Eq.(A8). This relation implies that a product with $g \leq 0$ will disappear from the market on the long time scale, because there are not sufficient financial reserves for investments. This is a well-known economic fact. The result also suggests that $g > 0$ for all surviving products. Note that a profit maximizing business unit will have a short life even if the profit per unit is positive, if the profit is not reinvested.

**Appendix B**

**The Price Distribution**

The corresponding Fokker-Planck equation of the Langevin equation (67) for the price probability distribution can be written as:

$$\frac{\partial P_\mu(\Delta\mu,t)}{\partial t} \sim \frac{\partial}{\partial\mu}\left(-F(\Delta\mu)P_\mu(\Delta\mu,t) + \frac{1}{2}D\frac{\partial P_\mu(\Delta\mu,t)}{\partial\mu}\right)$$

(B1)

For a sufficient long time the probability distribution approaches its stationary state:

$$P_\mu(\Delta\mu) \sim \exp\left(-\frac{2}{D}\Omega(\Delta\mu)\right)$$

(B2)

with the generalized potential

$$\Omega(\Delta\mu) = -\int F(\Delta\mu')d\Delta\mu'$$

(B3)

The integration can be carried out. With Eq.(65) we obtain for the stationary price distribution:

$$P_\mu(\Delta\mu) = \frac{b}{D}\exp\left(-\frac{2b}{D}|\Delta\mu|\right)$$

(B4)

The price distribution is therefore given by a Laplace (double exponential) distribution. The variance of the price distribution is time independent and given by:

$$Var(P_\mu) = \frac{1}{2}\frac{D^2}{b^2}$$

(B5)

**Appendix C**

We want to determine the growth rate distribution of the business units. Empirical studies investigate usually the growth rate difference of the business units for a given time interval $\Delta\tau$. Considering a time interval such that the replicator equation (Eq.(46)) is valid, the differential equation can be approximated by a difference equation:

$$\frac{y(\tau+\Delta\tau)-y(\tau)}{\Delta\tau}=ry(\tau)$$
(C1)

where $\tau >> \Delta\tau$. Scaling the time such that $\Delta\tau=1$, we can approximate the logarithm of the sales by the growth rate according to:

$$\log\left(\frac{y(\tau+1)}{y(\tau)}\right)=\log(r+1)\cong r$$
(C2)

Because fluctuations of the price are directly related to variations of the grow rate (Eq.61), the growth rate distribution of the unit sales can be obtained from the price distribution by changing variables. With:

$$r\cong\frac{df(\mu)}{d\mu}\Delta\mu$$
(C3)

we obtain for the stationary distribution:

$$P_r(r)\sim\exp(-|r|)$$
(C4)

Hence, the growth rate distribution is suggested to be also a Laplace distribution.

**Appendix D**

**The Evolution of the Mean Price for Durable Goods**

From Eq. (58) follows for the fitness at mean price:

$$f(\langle\mu\rangle) = \langle f \rangle = v(\langle\mu\rangle)\frac{\sum y_i \eta_i \gamma_i \psi_0}{y_t} = v(\mu)\langle\eta\gamma\psi_0\rangle$$

(D1)

where the brackets indicate the average over the sold products. Expanding the market volume *v(<μ>)* around $\mu_m$ up to the second order we obtain:

$$v(\langle\mu\rangle) \cong m_L\left(1 - \frac{(\langle\mu\rangle - \mu_m)^2}{2\Theta^2}\right)$$

(D2)

where we used Eq.(9). With this approximation we get:

$$\frac{df(\langle\mu\rangle)}{d\mu} = -\frac{\langle\gamma\eta\psi_0\rangle m_L}{\Theta^2}\left(\langle\mu\rangle - \mu_m\right)$$

(D3)

Hence, the price evolution is governed close to $\mu_m$ by:

$$\frac{d\langle\mu\rangle}{d\tau} = -\frac{\langle\gamma\eta\psi_0\rangle m_L Var(P_\mu)}{\Theta^2}\left(\langle\mu\rangle - \mu_m\right)$$

(D4)

For the case that the pre-factor on the right hand side can be considered as time independent, the integration can be carried out on the long time scale:

$$\mu(t) = \mu_0 e^{-at} + \mu_m$$

(D5)

where $\mu_0$ is an integration constant and the parameter

$$a = \frac{\varepsilon\langle\gamma\eta\psi_0\rangle m_L Var(P_\mu)}{\Theta^2}$$

(D6)

is denoted in this model as the price decline rate.

Whether the condition of a constant price decline rate is satisfied can be tested by a comparison with empirical data. The model suggests that the variance of the price distribution is a constant for uncorrelated price fluctuations. In this case the price decline rate can be further specified as:

$$a = \frac{\varepsilon\langle\gamma\eta\psi_0\rangle m_L D^2}{2\Theta^2 b^2}$$

(D7)

Note that a large price variance increases the price decline rate. This effect is a consequence of the selection process and is known as Fishers fundamental theorem of natural selection [48].

**References**

[1] M. A. Nowak, *Evolutionary Dynamics,* Harvard University Press (2006).

[2] R. Feistel, W. Ebeling, *Evolution of Complex Systems, Selforganization, Entropy and Development,* Kluwer Dordrecht (1989)

[3] E.D. Beinhocker, *The Origin of Wealth,* Harvard Business School Press (2006) p.14.

[4] E. M. Rogers, *Diffusion of Innovations,* eds. Simon and Schuster, New York, (2003).

[5] F. M. Bass, A new product growth model for consumer durables, *Management Science* 15 (1969) 215-227.

[6] B. Gompertz, On the Nature of the Function Expressive of the Law of Human Mortality, and on a New Mode of Determining the Value of Life Contingencies, *Philosophical Transactions of the Royal Society of London*. (1825) 513-583.

[7] N. Maede, and T. Islam, Modelling and forecasting the diffusion of innovation – A 25-year review, *International Journal of Forecasting* 22(3), 519-545, (2006).

[8] C. A. Silva, V. M. Yakovenko, Temporal evolution of the "thermal" and "superthermal" income classes in the USA during 1983–2001, Europhys. Lett. 69 (2005) 304-310.

[9] A. Banerjee, V.M. Yakovenko, T. Di Matteo, A study of the personal income distribution in Australia, Physica A 370 (2006) 54–59.

[10] V.M. Yakovenko, Econophysics, Statistical Approach to, in *Encyclopedia of Complexity and System Science*, edited by R. A. Meyers, Springer (2009)

[11] V.M. Yakovenko, J. B. Rosser, Colloquium: Statistical Mechanics of Money, Wealth, and Income, Reviews of Modern Physics 81, (2009) 1703-1725

[12] P.R. Steffens, A Model of Multiple Ownership as a Diffusion Process. Technological Forecasting and Social Change 70 (2002) 901–917.

[13] P.R. Steffens, An Aggregate Sales Model for Consumer Durables Incorporating a Time Varying Mean Replacement Age, J. Forecasting 20 (2001) 63-77.

[14] H.R. Varian, *Intermediate Microeconomics – A Modern Approach* W.W. Northern & Company Inc. New York (2006)

[15] R. Dawkings, The Blind Watchmaker, Penguin Books, (2006)

[16] J.C. Fisher, H.R. Pry, A simple substitution model of technological change, *General Electric Report No.70-C-215* (1970)

[17] J.L. McCauley, *Dynamics of Markets,* Cambridge University Press (2004)

[18] R. Feistel, W. Ebeling, Models of Darwin Processes and Evolution Principles, BioSystems 15 (1982) 291-299

[19] W. Ebeling, A. Engel, B. Esser, R. Feistel, Diffusion and Reaction in Random Media and Models of Evolutionary Processes, J. Stat. Phys. 37 (1984) 369-379

[20] M.H.R. Stanley, L.A.N. Amaral, S.V. Buldyrev, S. Halvin, H. Leshhorn, P. Maass, M.A: Salinger, H. E. Stanley, Scaling behaviour in the growth of companies, Nature 379 (1996) 804-806

[21] L.A. N. Amaral, S. V. Buldyrev, S. Havlin, M. A. Salinger, H. E.

Stanley, Power Law Scaling for a System of Interacting Units with Complex Internal Structure, Phys. Rev. Lett. 80 (1998) 1385-1388
[22] S.V. Buldyrev, J. Growiec, F. Pammolli, M. Riccaboni, H. E. Stanley The Growth of Business Firms: Facts and Theory, Journal of the European Economic Association, 5 (2007) 574–584.
[23] K. Matia, D. Fu, S. V. Buldyrev, F. Pammolli, M. Riccaboni and H. E. Stanley, Statistical properties of business firms structure and growth, Europhys. Lett., 67 (2004) 498–503
[24] G. De Fabritiis, F. Pamolli, M. Riccaboni, On size and growth of business units, Physica A 324.1-2 (2003) 38-44
[25] R. Gibrat, *Les Inegalites Economiques*, Sirey, Paris (1913)
[26] J. Growiec, F. Pammolli, M. Riccaboni, H.E. Stanley, On the size distribution of business firms, Economic Letters 98 (2008) 207-212
[27] G. Bottazzi, S. Sapio and A. Secchi, Some Statistical Investigations on the Nature and Dynamics of Electricity Prices. (2004) LEM Papers Series. RePEc:ssa:lemwps:2004/13.
[28] L.K. Vanston, R.L. Hodges, Technology Forecasting for Telecommunications, *Telektronikk* 4, 32-42 (2004)
[29] L.K. Vanston, Practical Tips for Forecasting New Technology Adoption, *Telektronikk* 3/4, 179-189 (2008)
[30] N. Economides, C. Himmelberg, Critical Mass and Network Size with Application to the US Fax Market. *Working Paper Series: Stern School of Business, NYU EC; 95-11 (1995)*
[31] Z. Wang, Income Distribution, Market Size and the Evolution of Industry, *Review of Economic Dynamics*, 11(3), 542-565, (2007).
[32] P.R. Steffens, The Product Life Cycle Concept: Buried or Resurrected by the Diffusion Literature? *Academy of Management Conference*, Technology and Innovation Management Division, Denver, (2002).
[33] C. Marchetti, Primary Energy Substitution Models: On the Integration between Energy and Society, Technological Forecasting and Social Change, 10 (1977) 345-356
[34] P.S. Meyer, J.W. Yung, J.H. Ausubel, A Primer on Logistic Growth and Substitution: Mathematics of the Loglet Lab Software, Technological Forecasting and Social Change, 61 (1999) 247-271
[35] V. Mahajan, E. Muller, Timing, Diffusion, and Substitution of Successive Generations of Technological Innovations: The IBM Mainframe Case, Technological Forecasting and Social Change, 51 (1996) 109-132
[36] N.M. Victor, J.H. Ausubel, DRAMs as Model Organisms for Study of Technological Evolution, Technological Forecasting and Social Change, 69 (2002) 243-262
[37] S. Park, Strategic manoeuvring and standardization in the presence of network externalities: a simulation study of the VCR case, IEEE (2003) 205-214
[38] H. Ohashi, The Role of Network Effects in the US VCR Market, 1978-1986, J. of Economics & Management Strategy, 12 (2003) 447–494
[39] M.L. Katz, C. Shapiro Product Introduction with Network Externalities. J. Industrial Economics 40 (1992) 55-84.
[40] C. Shapiro, H.R. Varian, Network effects. Notes to Accompany Information Rules: A Strategic Guide to the Network Economy. Harvard Business School Press (1998).

[41] F.M. Bass, T.V. Krishnan, D. Jain, Why the Bass Model fits without Decision Variables, Marketing Sci. 13 (1994) 203-223.

[42] V. Mahajan, E. Muller, Y. Wind, *New- Product Diffusion Models*, Springer (2000).

[43] D. Chandrasekaran, G. J. Tellis, A Critical Review of Marketing Research on Diffusion of New Products, Rev. Marketing Research. 39-80, (2007).

[44] L. Tvede, Business Cycles, History, Theory and Investment Reality, Jon Wiley & Sons (2006). p 169

[45] J. A. Schumpeter, *The theory of economic development : An Inquiry into Profits, Capital, Credit, Interest and the Business Cycle*. Cambridge MA, Harvard University Press (1989).

[46] R.R. Nelson, S.G. Winter, *A Evolutionary Theory of Economic Change*. Harvard University Press MA, (1982).

[47] K. Dopfer, *The Evolutionary Foundation of Economics*. University Press Cambridge UK, (2005).

[48] R.A. Fisher, *The genetical theory of natural selection*. Clarendon, Oxford (1930).

**Tables**

| *Parameter* | *Colour TV* | *FAX* | *B&W TV* | *Clothes Dryer* | *Air Conditioner* | *VCR* |
|---|---|---|---|---|---|---|
| $t_0$ | 1954 | 1977 | 1948 | 1949 | 1951 | 1976 |
| $\Delta t$[years] | 1/2 | 4 | 0 | 0 | 0 | 0 |
| $p_m/p_0$ | 0 | 0.01 | 0.33 | - | - | 0.17 |
| $a$[1/year] | 0.103 | 0.45 | 0.2 | 0.081 | 0.11 | 0.195 |
| $k$ | 27 | 360 | 8.5 | 25 | 65 | 55 |
| $n_{G0}$ | 0.97 | 0.98 | 0.77 | 0.9 | 0.9 | 0.83 |
| $n_{B0}$ | 0.01 | 0.02 | 0.18 | 0.1 | 0.1 | 0.03 |
| $A$ | 0.001 | 0.01 | 0.02 | 0.02 | 0.03 | 0.01 |
| $B$ | 1.8 | 2.2 | 2.5 | 1 | 0.8 | 1 |
| $R$ | - | - | 0.3 | 0 | 0 | - |
| $Q$ | - | 2.5 | 0.06 | 0.06 | 0.02 | 0.3 |
| $R'$ | - | - | 0.65 | 0 | 0 | - |
| $Q'$ | - | 0 | 0.06 | 0.35 | 0.33 | 0 |
| $t_p$[years] | - | - | 9.2 | - | - | - |
| $t'_p$[years] | - | - | 10.2 | - | - | - |
| $M\ 10^6$ | - | - | ~53 | ~35 | ~45 | ~96 |

**Table 1.** Characteristic parameters of the studied examples.

**Figures**

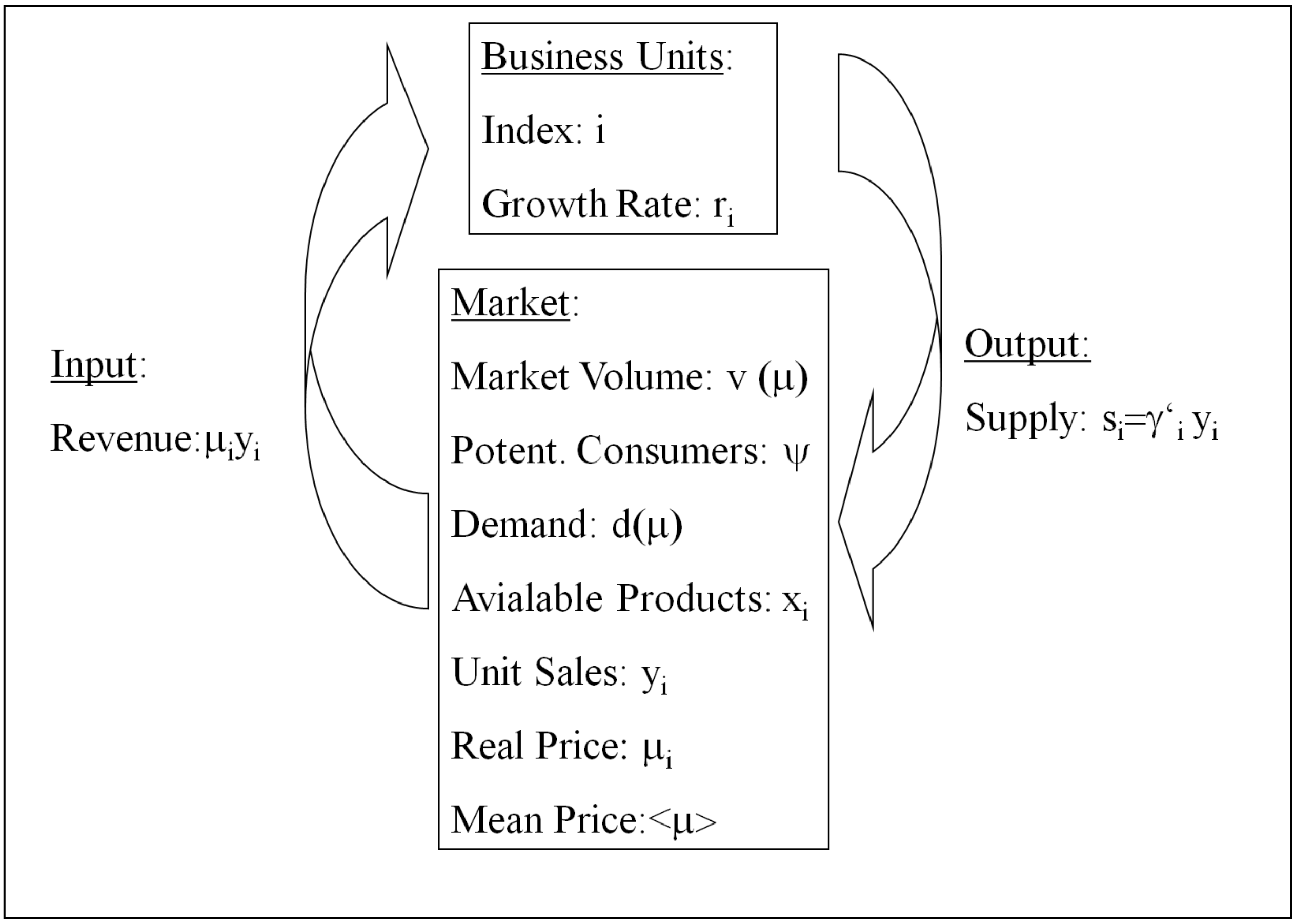

**Figure 1:** The reproduction process.

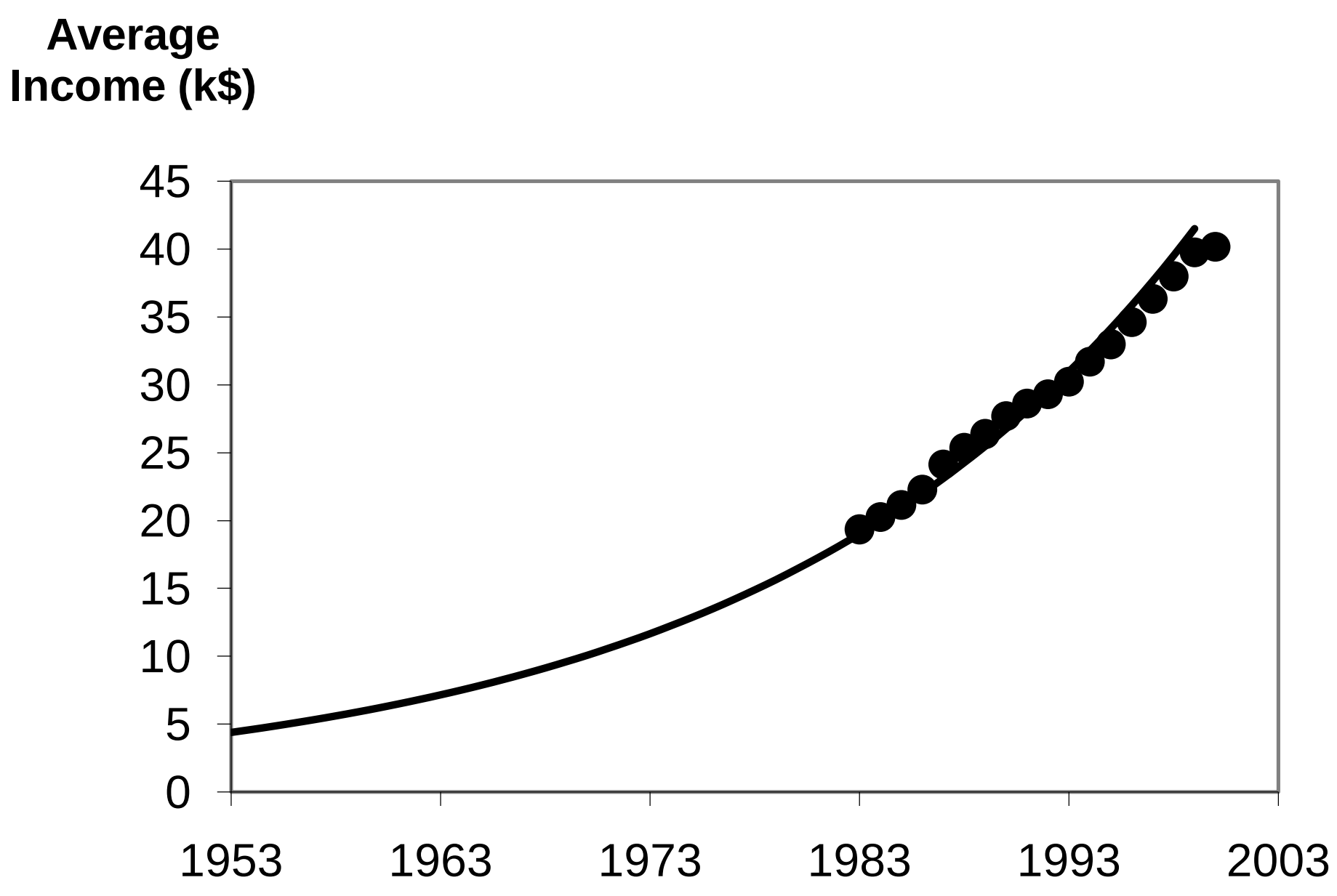

**Figure 2:** Average income of the lower class in the USA (circles). The solid line displays Eq.(94).

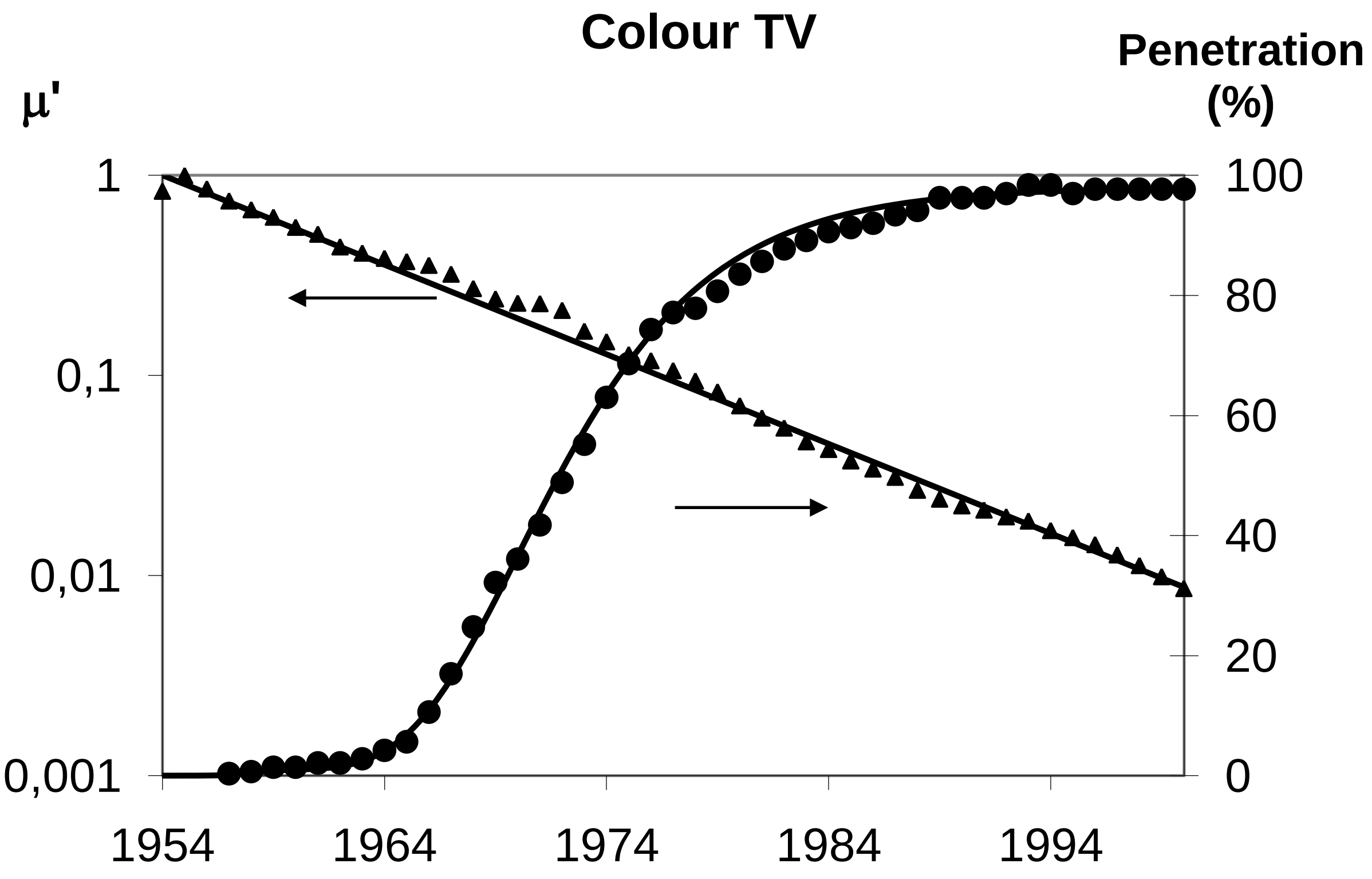

**Figure 3:** Evolution of the price $\mu'$(triangles) and market penetration (dots) of Colour TV sets in the USA [29].

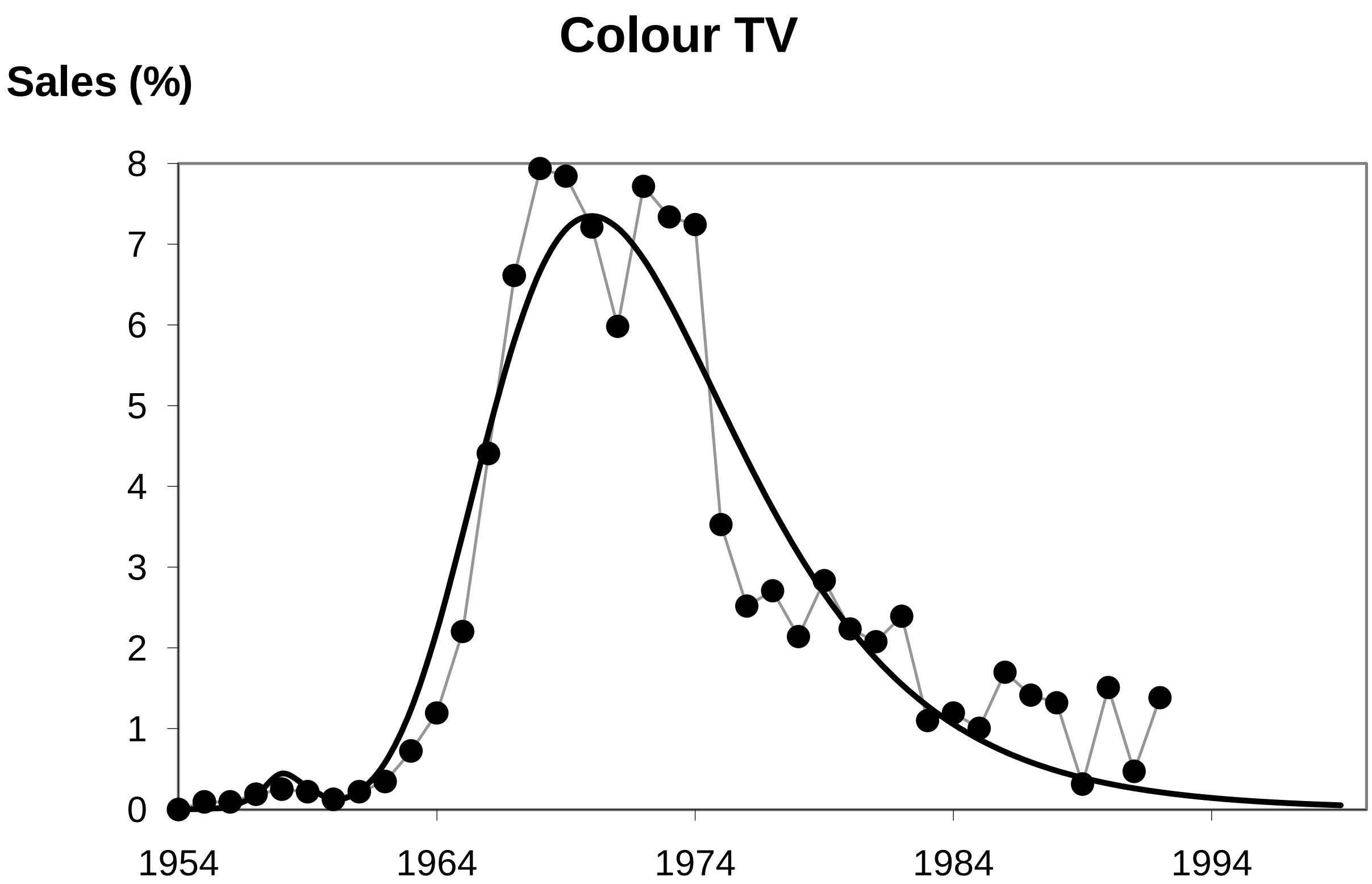

**Figure 4:** First purchase sales of Colour TV sets in the USA in percentage of households [28]. The solid line follows directly from Eq.(27) with the parameters in Table 1.

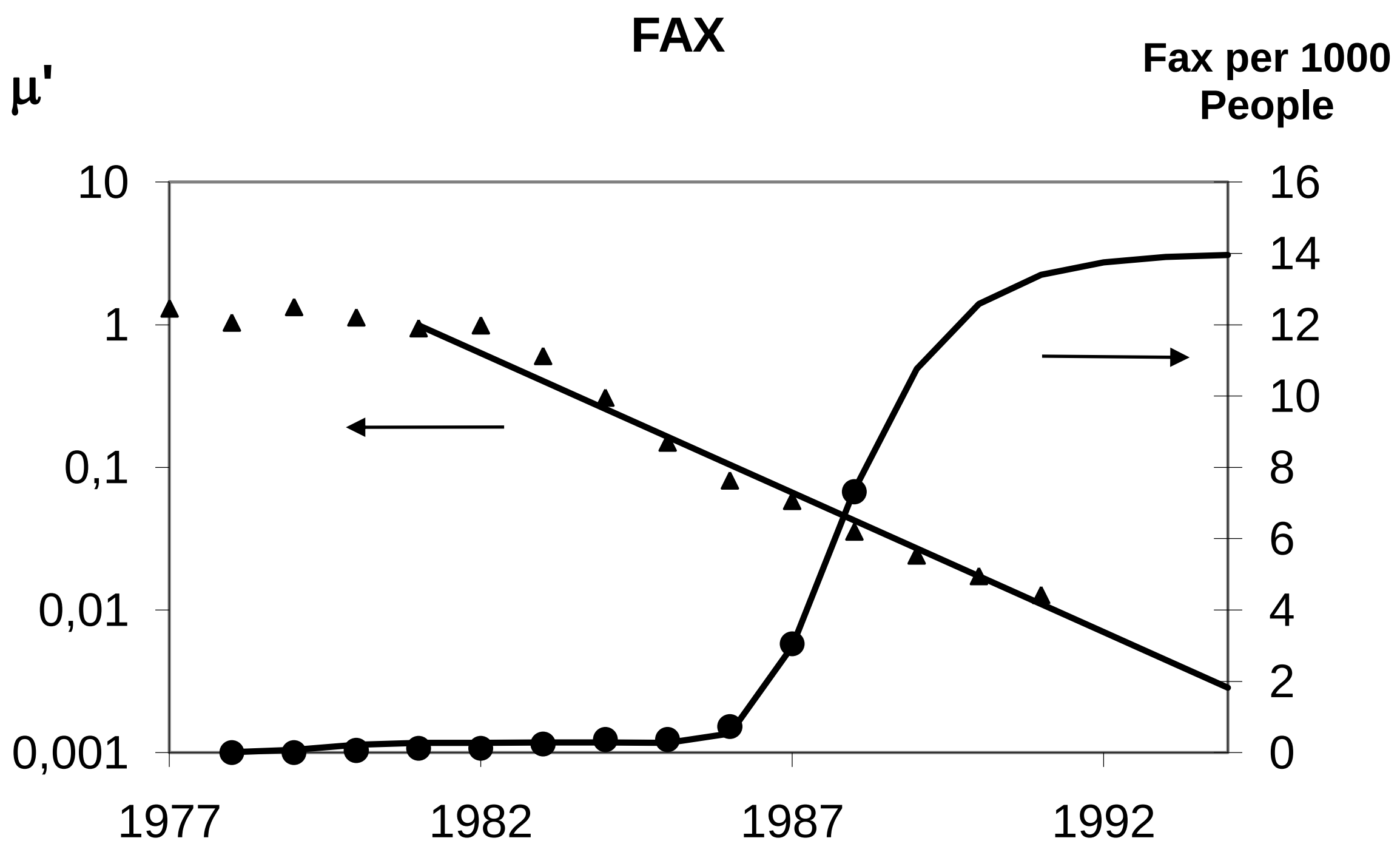

**Figure 5:** Evolution of the price function $\mu'$ (triangles) and market penetration (dots) of Fax machines in the USA [30].

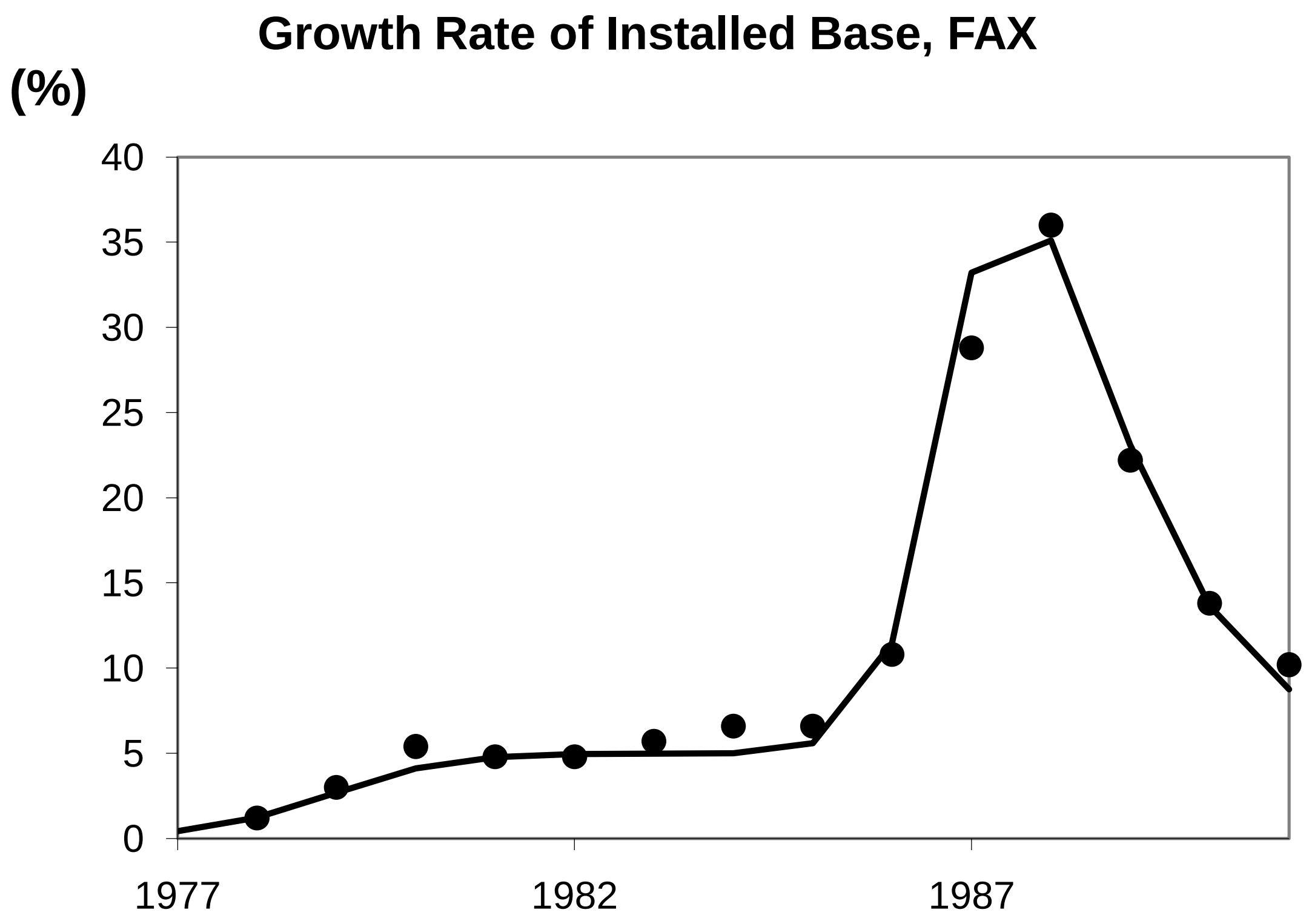

**Figure 6:** Evolution of the installed base of Fax machines in the USA [30].

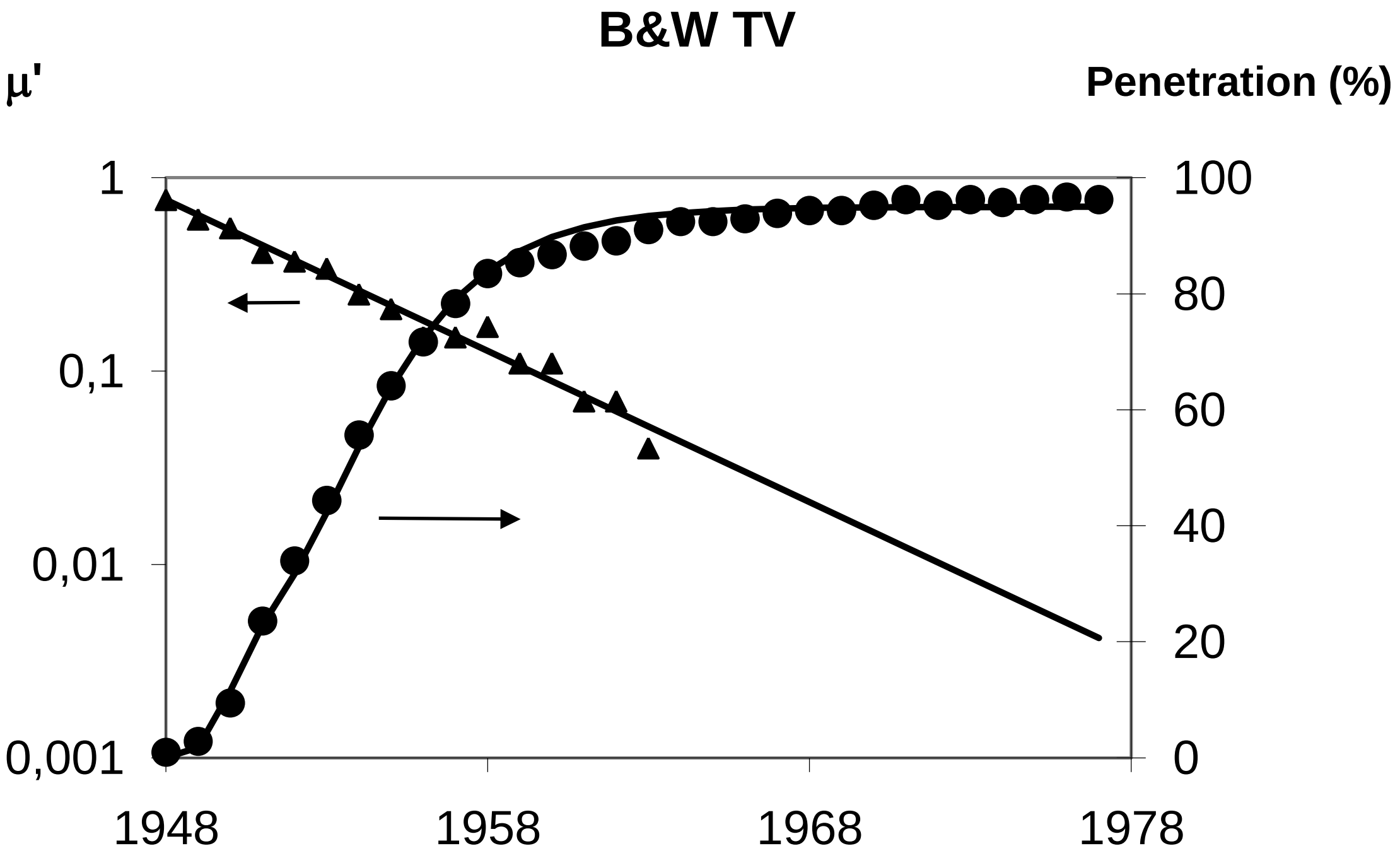

**Figure 7:** Evolution of the price function $\mu'$ (triangles) and market penetration (dots) of Black & White TV sets in the USA [31].

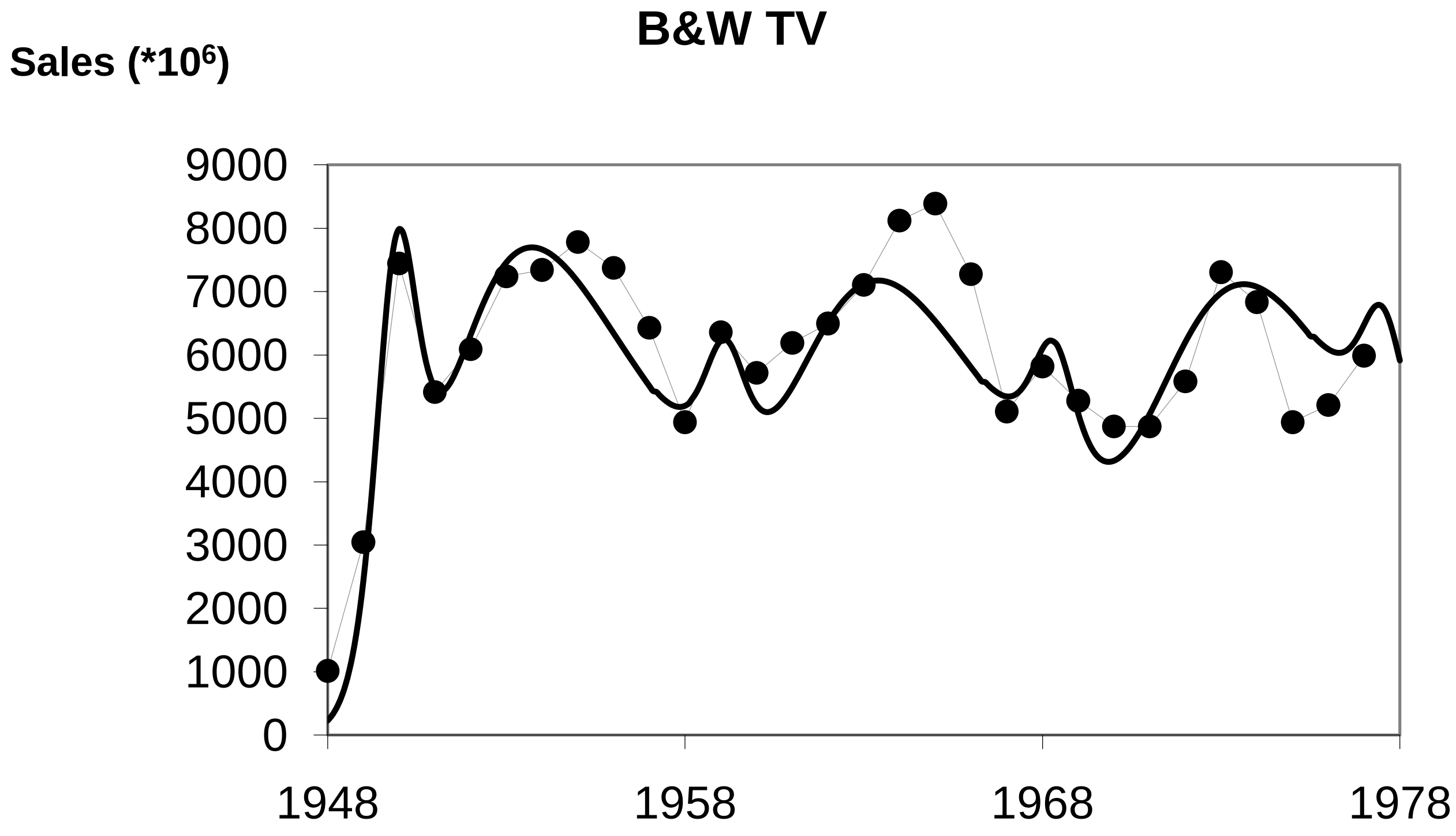

**Figure 8:** Evolution of the unit sales of Black & White TV sets in the USA [32]. The fat line is a fit of the product life cycle. The periodicity of the sales is caused by replacement purchase. The narrow peaks are linked to Bass diffusion and the broad peaks to Gompertz diffusion.

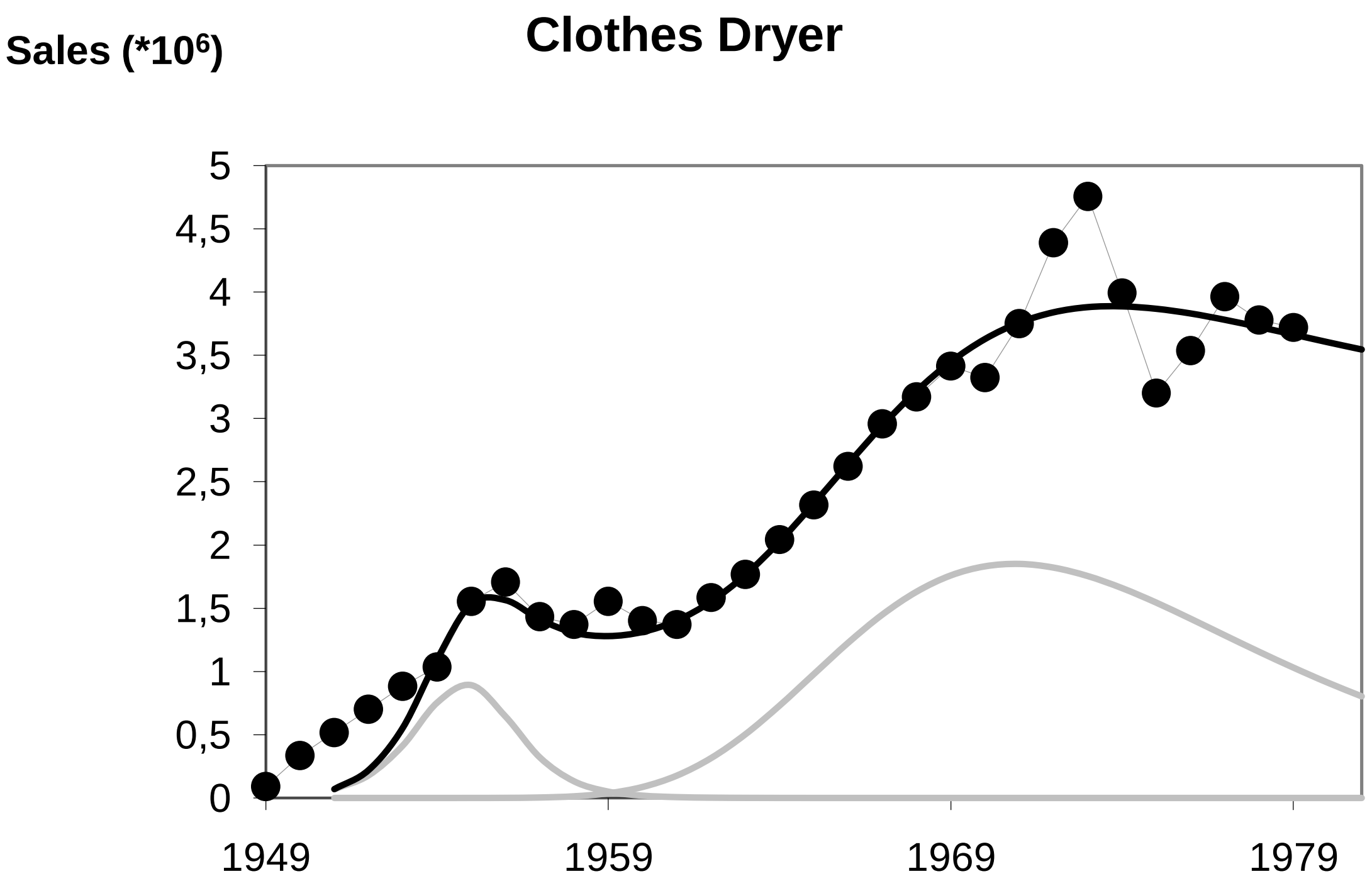

**Figure 9:** Evolution of the unit sales of clothes dryers in the USA [32]. The grey lines represent first purchase sales, while the first peak corresponds to Bass diffusion, and the second to Gompertz diffusion.

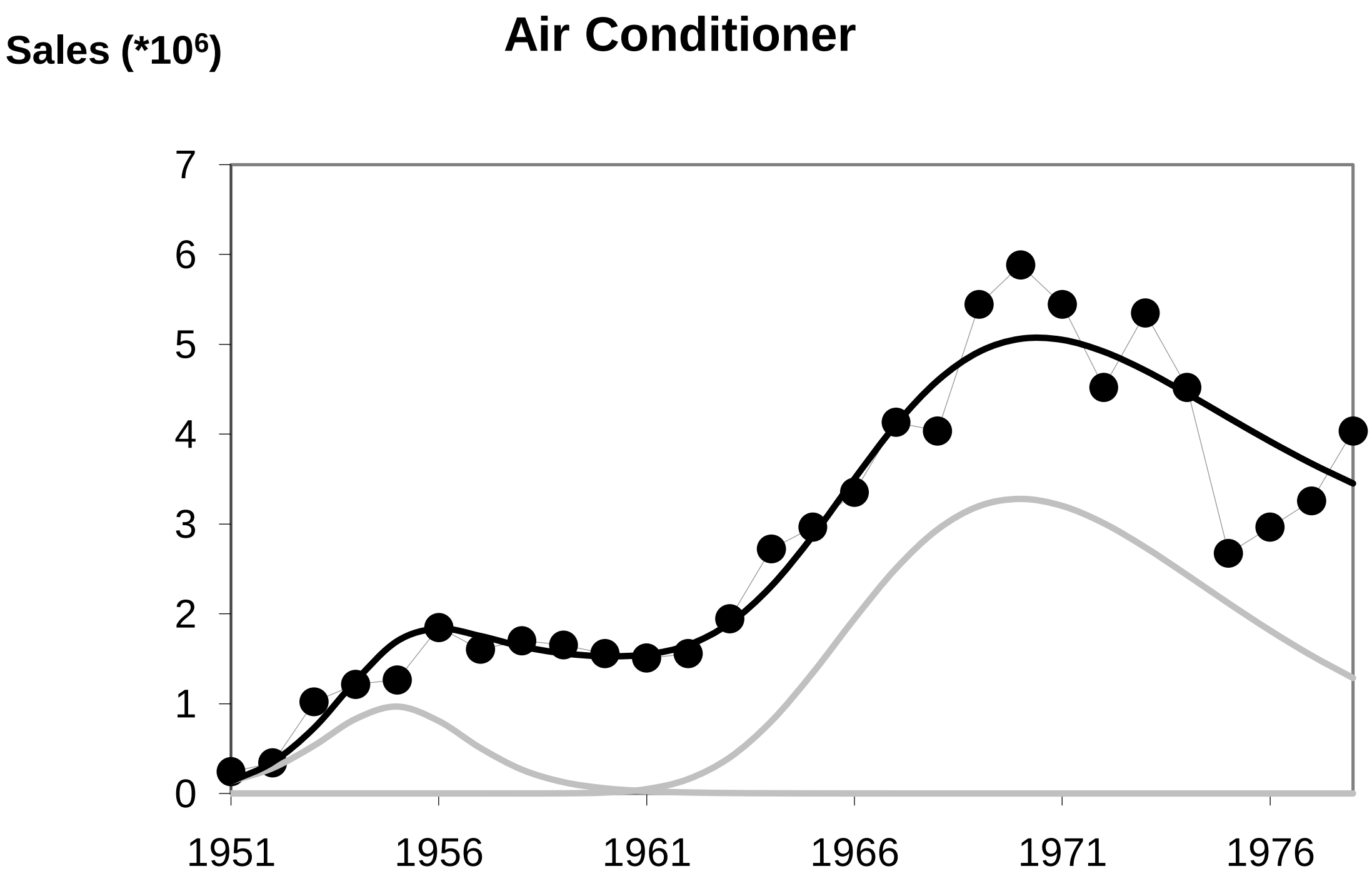

**Figure 10:** Evolution of the unit sales (dots) of air conditioners in the USA [32]. The grey lines represent first purchase sales, while the first peak corresponds to Bass diffusion, and the second to Gompertz diffusion.

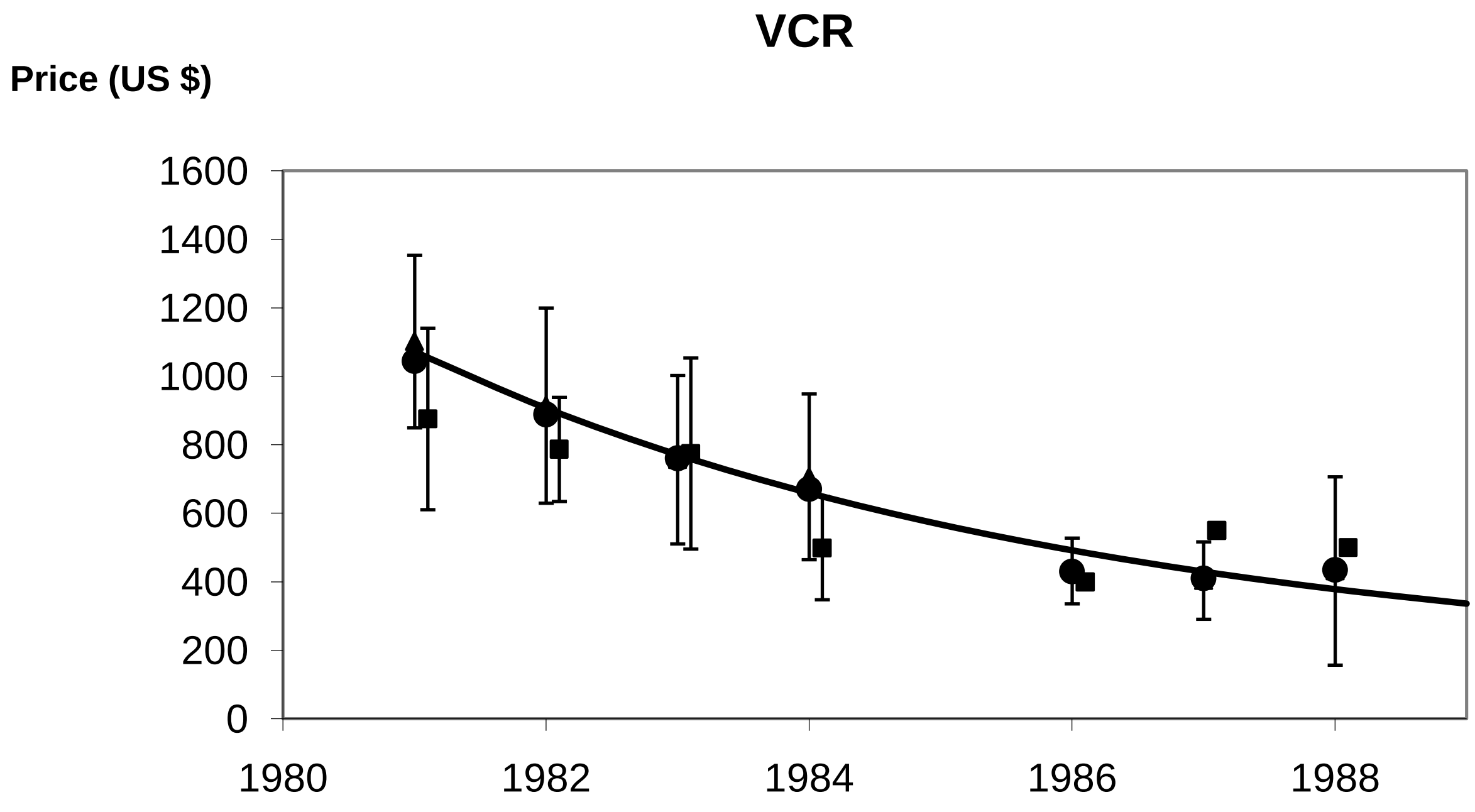

**Figure 11:** Evolution of the mean price (dots) and the average price for VHS (triangles) and Betamax (squares) VCR systems in the USA [37]. The error bars indicate the standard deviation of the price. For a better identification the Betamax data are slightly shifted to the right.

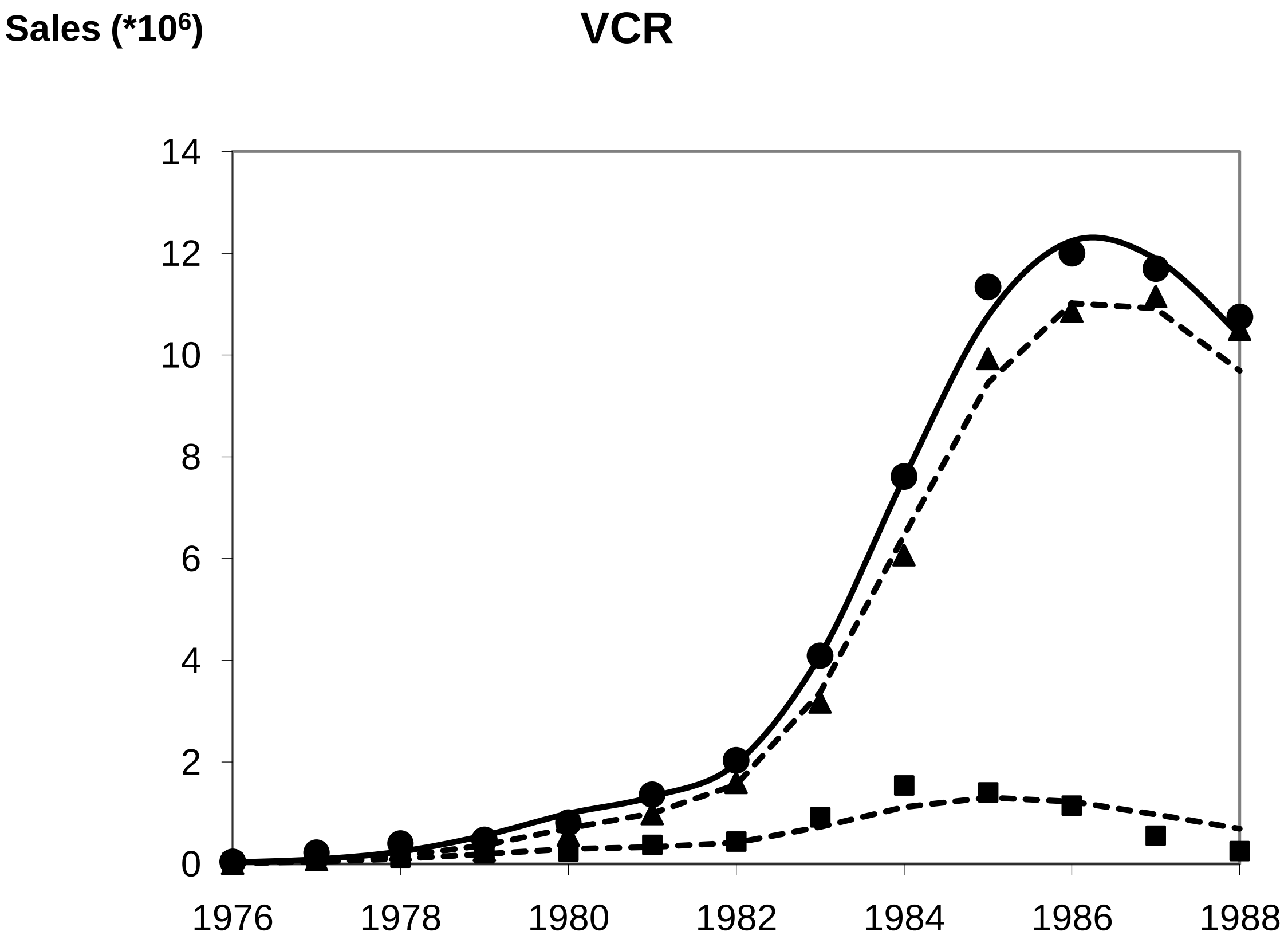

**Figure 12:** Evolution of the unit sales (dots), for VHS (triangles) and Betamax (squares) VCR- systems in the USA [37]. The solid line is a fit of the total sales, while the dotted lines are the product of the predicted market share (Fig.11) with the total sales.

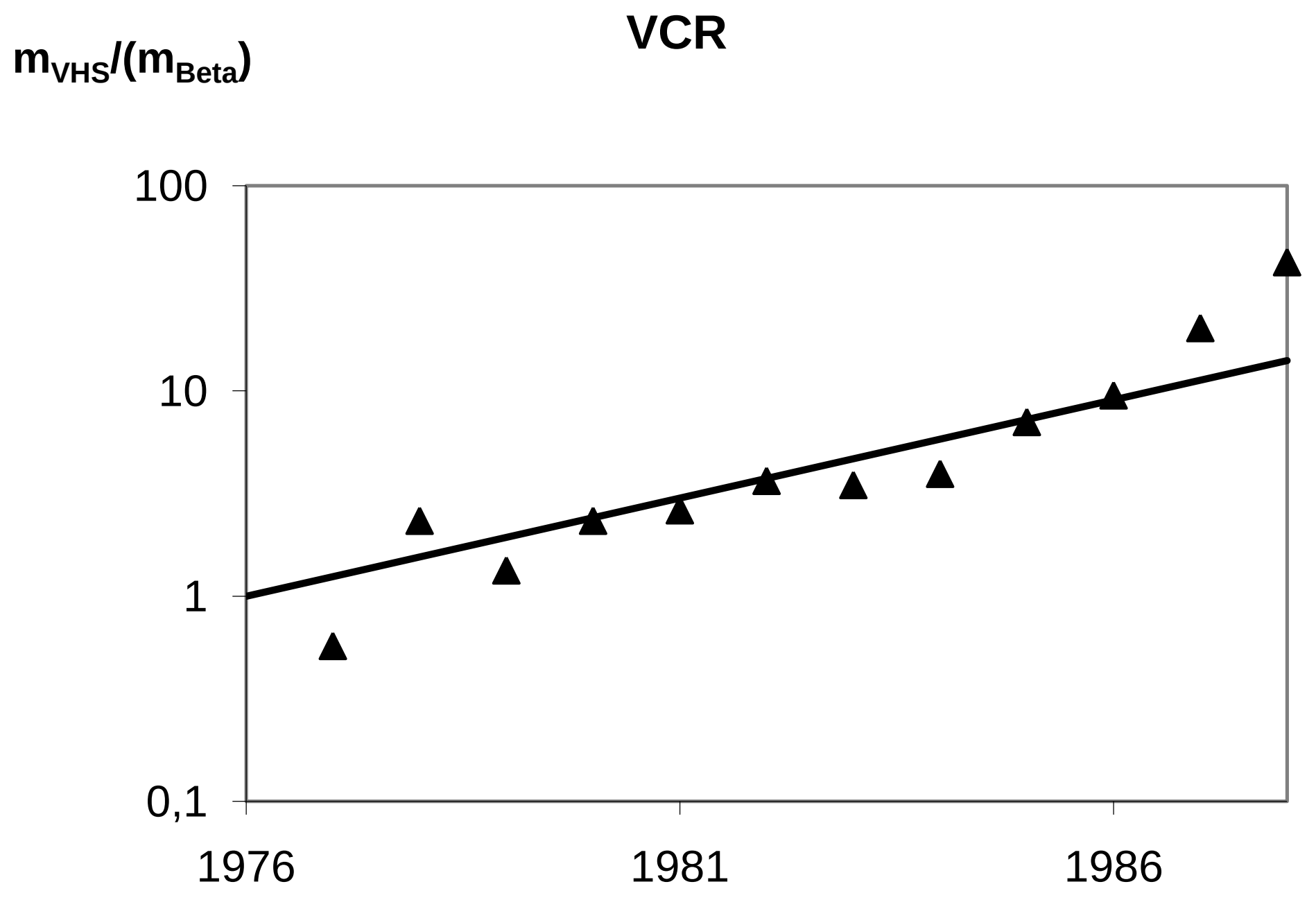

**Figure 13:** Evolution of the VHS sales market share in the USA [37].

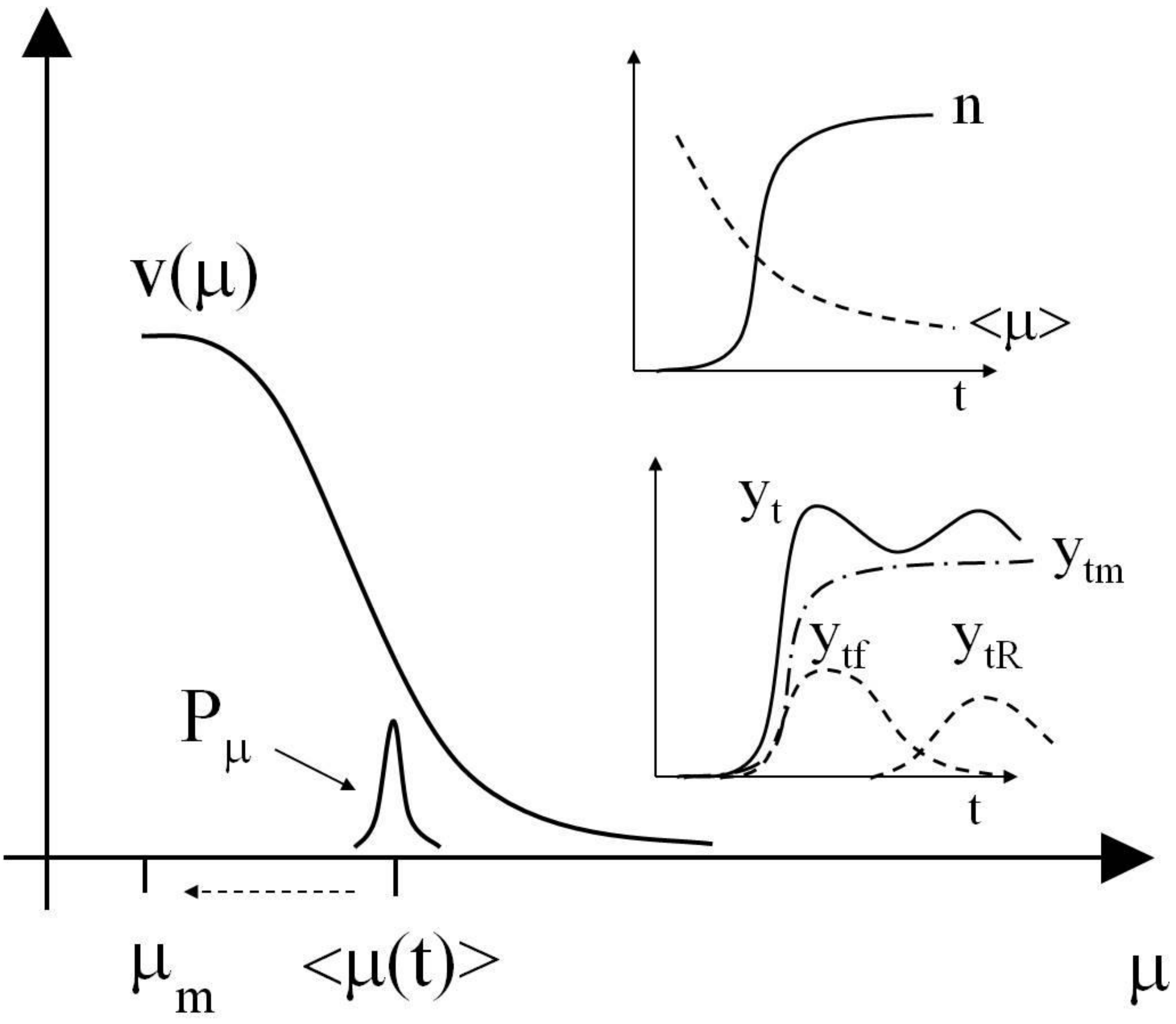

**Figure 14:** Schematically displayed is the market evolution of a durable good neglecting Bass diffusion (see text).